\documentclass[prb,twocolumn]{revtex4-1}

\usepackage{amsfonts}
\usepackage{amsmath}
\usepackage{amssymb}
\usepackage{graphicx}
\usepackage[T1]{fontenc}
\usepackage[utf8]{inputenc}
\usepackage{hyperref}
\usepackage{xcolor}
\usepackage{color}
\usepackage{bm}
\usepackage{times}
\usepackage{comment}
\usepackage{footnote}
\DeclareMathAlphabet      {\mathbfit}{OML}{cmm}{b}{it}
\newcommand{\BE}{\begin{equation}}
\newcommand{\BEA}{\begin{eqnarray}}
\newcommand{\EE}{\end{equation}}
\newcommand{\EEA}{\end{eqnarray}}

\newcommand\R\rangle
\renewcommand\L\langle
\renewcommand\r\right
\renewcommand\l\left

\definecolor{vertforet}{rgb}{0.202496,0.606335,0.181094}

\hypersetup{
    bookmarks=true,         			% show bookmarks bar?
    unicode=false,         			 % non-Latin characters in Acrobatâ
    pdftoolbar=true,        			% show Acrobat
    pdfmenubar=true,        			% show Acrobat
    pdffitwindow=false,     			% window fit to page when opened
    pdfstartview={FitH},    			% fits the width of the page to the window
    pdftitle={My title},    			% title
    pdfauthor={Author},     			% author
    pdfsubject={Subject},   			% subject of the document
    pdfcreator={Creator},   			% creator of the document
    pdfproducer={Producer}, 			% producer of the document
    pdfkeywords={keyword1} {key2} {key3}, 	% list of keywords
    pdfnewwindow=true,      			% links in new window
    colorlinks=true,       			% false: boxed links; true: colored links
    linkcolor=vertforet,          		% color of internal links
    citecolor=blue,       			% color of links to bibliography
    filecolor=magenta,      			% color of file links
    urlcolor=cyan           			% color of external links
}

\newcommand{\eq}[1]{\begin{align} #1 \end{align}}
\renewcommand{\H}{\mathcal{H}}

\newcommand{\pa}[1]{\left( #1 \right)} 
\newcommand{\ac}[1]{\left\{ #1 \right\} }

\newcommand{\av}[1]{\left< #1 \right>}

\renewcommand{\H}{\mathcal{H}}
\newcommand{\Hm}{\mathcal{H}_\mathrm{MF}}
\newcommand{\jzz}{J_{zz}}
\newcommand{\jp}{J_{\pm}}
\renewcommand{\jpp}{J_{\pm\pm}}
\newcommand{\jzp}{J_{z\pm}}
\newcommand{\jszz}{j_{zz}}
\newcommand{\jspp}{j_{\pm\pm}}
\newcommand{\jszp}{j_{z\pm}}

\newcommand{\Sl}{S}

\begin{document}
\title{Order-by-Disorder Near Criticality in {\bf{\it XY}} Pyrochlore Magnets}

\author{Behnam Javanparast}\affiliation{Department of Physics and
  Astronomy, University of Waterloo, Waterloo, ON, N2L 3G1, Canada}
\author{Alexander G. R. Day}\affiliation{Department of Physics and
  Astronomy, University of Waterloo, Waterloo, ON, N2L 3G1, Canada}
\author{Zhihao Hao}\affiliation{Department of Physics and
  Astronomy, University of Waterloo, Waterloo, ON, N2L 3G1, Canada}
\author{Michel J. P.  Gingras}\affiliation{Department of Physics and
  Astronomy, University of Waterloo, Waterloo, ON, N2L 3G1, Canada}
\affiliation{Canadian Institute for Advanced Research, 180 Dundas
  St. W., Toronto, ON, M5G 1Z8, Canada} \affiliation{Perimeter
  Institute for Theoretical Physics, 31 Caroline St. N., Waterloo, ON,
  N2L 2Y5, Canada}

\begin{abstract}
We consider a system of spins on the sites of a three-dimensional pyrochlore lattice of corner-sharing tetrahedra 
interacting with a predominant effective $xy$ exchange. 
In particular, we investigate the selection of a long-range ordered state with broken discrete symmetry induced by thermal fluctuations
near the critical region.
At the standard mean-field theory (s-MFT) level,  in a region of the parameter space of this Hamiltonian that 
we refer to as $\Gamma_5$ {\it region}, the ordered state possesses an accidental $U(1)$ degeneracy. In this paper, we show that fluctuations beyond s-MFT lift this degeneracy by selecting one of two states (so-called $\psi_2$  and  $\psi_3$) from the degenerate manifold, thus
exposing a certain form of order-by-disorder (ObD). We analytically explore this selection at the microscopic level and close to criticality by elaborating upon and using an extension of the so-called TAP method, originally developed by Thouless, Anderson and Palmer to study the effect of fluctuations in spin glasses.  
We also use a single-tetrahedron cluster-mean-field theory (c-MFT) to explore over what minimal length scale fluctuations can lift the degeneracy.
 We find the phase diagrams obtained by these two methods to be somewhat different since c-MFT only includes the shortest-range fluctuations.
General symmetry arguments used to construct a Ginzburg-Landau theory to lowest order in the order parameters predict that a weak magnetic moment, $m_z$, 
along the local $\langle 111 \rangle$ (${\hat z}$) direction is generically induced for a system ordering into a $\psi_2$ state, but not so for $\psi_3$ ordering.
Both E-TAP and c-MFT calculations confirm this weak fluctuation-induced $m_z$ moment.
Using a Ginzburg-Landau theory, we discuss the phenomenology of multiple phase transitions below the paramagnetic phase transition and within
the $\Gamma_5$ long-range ordered phase.
%that has heretofore been largely ignored in theoretical studies of $xy$ pyrochlore magnets. 

\end{abstract}

%\pacs{71.10.Fd, 71.30.+h, 75.10.Jm, 75.40.Mg, 75.50.Ee}

%\date{today}

\maketitle

\section{Introduction} 

In the study of condensed matter systems, mean-field theory \cite{cond} is often the simplest starting point to obtain a qualitative understanding of the essential physics at play prior to carrying out a more sophisticated analysis.  A standard mean-field theory (s-MFT) replaces the many-body problem with a simpler problem of a one-body system interacting with an averaged field produced by the rest of the interacting particles.
In systems with competing or frustrated interactions, s-MFT may yield a number of states with a degenerate minimum free energy below the mean-field critical temperature, $T_c^{\mathrm{MF}}$. \cite{Reimers,Matt2} 
If these degeneracies are accidental, that is not imposed by exact symmetries of the Hamiltonian,  they may be
%These accidental degeneracies, which are not imposed by exact symmetries of the Hamiltonian describing the interactions, are 
lifted by the effects of thermal\cite{villain1980order} or quantum\cite{shender_ObD} fluctuations,  a phenomenon known as 
order-by-disorder (ObD).\cite{villain1980order,shender_ObD,Henley_ObD,Shender_Peter,Yildirim_ObD}

The concept of ObD was originally proposed by Villain and collaborators as the 
ordered state selection mechanism for a two-dimensional frustrated Ising model on the domino 
lattice.\cite{villain1980order} Since this seminal work, ObD has been theoretically 
identified and discussed for many highly frustrated magnetic 
models.\cite{Henley_ObD,Yildirim_ObD,Shender_Peter,Sachdev-qObD,Moessner_Chalker, Ch_Hold_Shen, DipObD_Chen_Moess, Zhito_trian_impu,pawel1,Zhitomirsky:2012fk,Savary:2012uq,Wong,Oitma1} 
These systems generically possess an exponentially ($\exp[N^\alpha]$)
large number of classical ground-states (here $N$ is the number of spins in the system and $\alpha \le 1$).
 As a result, highly frustrated magnetic systems are intrinsically very  sensitive to fluctuations or energetic perturbations. 
In the context of experimental studies of real materials, it is difficult to distinguish a selection of an 
ordered state via fluctuations from one that would arise from
energetic perturbations beyond the set of interactions considered in a restricted theoretical model.
 Consequently, undisputed examples of ObD in  experiments have remained scarce.~\cite{garnet_ObD1,garnet_ObD2,SrCuOCl_ObD}

\begin{figure}[h]
\centering
\includegraphics[scale=0.75]{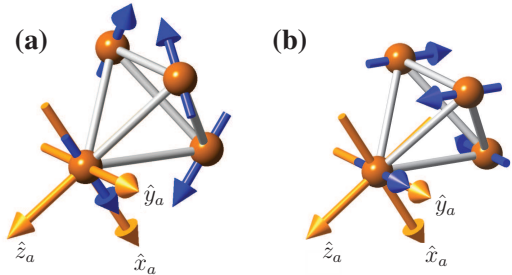}
%\put(-200,100){\large{\bf{(a)}}}
%\put(-100,100){\large{\bf{(b)}} }
\caption{(Color Online). Spin configurations of the {\bf (a)} non-coplanar $\psi_2$ state, as observed in
Er$_2$Ti$_2$O$_7$,~\cite{Poole:2007ys}
and the {\bf (b)} coplanar $\psi_3$ state on a single tetrahedron. The pattern is the same on all tetrahedra in a pyrochlore lattice which is a
face-centered cubic lattice with a tetrahedron basis
and these two states, which belong to the $\Gamma_5$ manifold, are thus said to have a ${\bm k}=0$ ordering wave vector.
The spin (in blue) at the $a$ sublattice points along 
{\bf (a)}
the $\hat x_a$ axis of the local $[111]$ reference frame
for the  $\psi_2$ state and along 
{\bf (b)} the $\hat y_a$ axis of the same local frame for the $\psi_3$ state.
The local $\hat z_a$  axis at sublattice $a$ points along the local cubic $[111]$ 
axis at that site such that the local $\hat x_a, \hat y_a, \hat z_a$ triad of orthogonal unit vectors (orange arrows) fulfills  
$\hat x_a \times \hat y_a =\hat z_a$.
As illustrated in panels {\bf  (a)} and {\bf (b)}, 
the local $\hat z_a$ axis for sublattice $a$ points directly out of the tetrahedron primitive basis cell.
See Appendix \ref{apPsi23} for more details.
}
\label{psi23}
\end{figure}

Quite recently, following an original proposal going back ten years,\cite{Champion_Er2Ti2O7, Champ_Peter_Psi2}
and building on an earlier study,\cite{BGR}
several papers \cite{Zhitomirsky:2012fk,Savary:2012uq,Wong,Oitma1} have put forward compelling arguments 
for ObD being responsible for the experimentally observed long-range order in the insulating rare-earth pyrochlore oxide~\cite{Gardner_RMP} 
Er$_2$Ti$_2$O$_7$. In this compound, Er$^{3+}$ is magnetic and Ti$^{4+}$ is not.
The key observation in those works is that the accidentally degenerate classical ground states are related by operations with 
a $U(1)$ symmetry. \cite{Savary:2012uq} This set of classically degenerate states form the 
so-called $\Gamma_5$ \emph{manifold}.\cite{Poole:2007ys} 
For a range of interaction parameters of the most general 
symmetry-allowed bilinear nearest-neighbor pseudospin $1/2$  exchange-like Hamiltonian (referred to as ${\cal H}$ in Eq.~(\ref{h}) below ) 
on the pyrochlore lattice of corner-sharing tetrahedra (see. Fig.~\ref{pyro}), the 
$U(1)$ degeneracy is exact at the s-MFT level as long as the cubic symmetry of the system remains intact. 
As a shorthand, we henceforth refer to this region of exchange parameter space as the $\Gamma_5$  \emph{region}, as referred to in the Abstract. 
%Based on the robustness of the $U(1)$ symmetry, Ref. [\onlinecite{Savary:2012uq}] argues 
%that essentially \emph{only} fluctuations can lift the degeneracy in Er$_2$Ti$_2$O$_7$.
Reference [\onlinecite{Savary:2012uq}] showed that the $U(1)$ symmetry is robust against a wide variety of perturbations added to ${\cal H}$ and
argued that essentially only fluctuations can efficiently
lift the degeneracy in Er$_2$Ti$_2$O$_7$ when considering bilinear anisotropic interactions 
of arbitrary range between the pseudospins.{\footnote{It has been pointed out by McClarty \emph{et al.} that 
the $\psi_2$ state can be energetically selected through a 
Van Vleck-like mechanism.~\protect{\cite{McClarty_Er2Ti2O7}}  
Ref.~[\onlinecite{Savary:2012uq}] argues that this mechanism is too weak 
by several orders of magnitude to be at play in Er$_2$Ti$_2$O$_7$.
However, a recent paper\protect{\cite{Petit_ObD}}
 questions this assertion.}}
Similar arguments were made in Ref.~[\onlinecite{Zhitomirsky:2012fk}].
In this compound, a particular long-range ordered state, the $\psi_2$ state 
(see Fig. (\ref{psi23})), \cite{Champion_Er2Ti2O7, Champ_Peter_Psi2, Poole:2007ys, Ruff_ETO, Dalmas_ETO,Ross_gap,Petit_ObD}
 is selected. Considering ${\cal H}$ on the pyrochlore lattice, Wong \emph{et al.} \cite{Wong} and Yan \emph{et al.}
 \cite{Ludo_edge} studied the effect of 
quantum \cite{Wong,Ludo_edge} and thermal \cite{Ludo_edge} 
fluctuations and established a general phase diagram for this Hamiltonian at $T=0^+$. 

The investigations reported in Refs.~[\onlinecite{pawel1,Zhitomirsky:2012fk,Savary:2012uq,Wong,Ludo_edge}] 
focused on identifying the mechanism of {\it ground state} selection
by taking into account the harmonic quantum~\cite{pawel1,Zhitomirsky:2012fk,Savary:2012uq,Wong,Ludo_edge}
or classical~\cite{pawel1,Ludo_edge} spin fluctuations about a classical long-range ordered state. 
On the other hand,  the problem of state selection at temperatures near the critical transition temperature 
to the paramagnetic phase has received significantly less attention. 
Although there have been numerical studies of ObD selection 
of the $\psi_2$ state at
 $T \lesssim T_c$,~\cite{Champion_Er2Ti2O7, Champ_Peter_Psi2, Ludo_edge, pawel1,Zhitomirsky:2012fk,Zhitomirsky_danger,Chern:2010vn} 
or upon approaching $T_c$ from above,~\cite{Oitma1} an analytical study specifying the role of the individual 
microscopic anisotropic spin-spin interactions in the ObD mechanism at $T\approx T_c$ has, 
to the best of our knowledge, not yet been carried out.
The problem of selection at $T\lesssim T_c$ is not only of relevance to the phenomenology of 
Er$_2$Ti$_2$O$_7$ or other pyrochlore magnetic compounds,~\cite{Gardner_RMP}
 but it is of considerable interest for all highly frustrated magnetic systems proposed to display an ObD mechanism.

There are situations where a sort of ObD occurs near $T_c$ which differ
from the textbook cases \cite{villain1980order,shender_ObD,Henley_ObD,Yildirim_ObD}
  where the state selection near $T=0^+$, proceeding either via thermal or quantum selection,
is  leveraged upon all the way to the long-range ordered state selected at $T_c$.
For example, the long-range ordered state selected by ObD can in principle be different for the $T=0^+$ and $T \lesssim T_c$ regimes. 
This occurs, for example, for classical Heisenberg spins on the pyrochlore lattice interacting 
via nearest-neighbor antiferromagnetic exchange and indirect  Dzyaloshinskii-Moriya interaction.~\cite{Canals_DM,Chern:2010vn}
In another class of problems, different competing long-range ordered states may have the same free-energy at the
s-MFT level only over a finite temperature interval, $T^* \le T \le T_c$ with a transition to 
a  \emph{non-degenerate} classical ground state at $T^*$. There, thermal fluctuation corrections to s-MFT can select one of the competing states
over the $T^* \le T \le T_c$ window. This is what is predicted to occur in the multiple-${\bm k}$ state selection in the
Gd$_2$Ti$_2$O$_7$ pyrochlore antiferromagnetic between 0.7 K and 1.0 K.~\cite{GTO-us}
Perhaps the most exotic cases arise when a relic of ObD occurs at the critical temperature, 
while the classical ground state is \emph{not} degenerate for the Hamiltonian considered, but would be for a closely related
Hamiltonian minus some degeneracy-lifting weak energetic perturbations.~\cite{pawel1,Ch-Tch-Moe,Pinettes-Canal} 
At the phenomenological Ginzburg-Landau level description, these cases are not paradoxical.
They only become so when one is trying to ascribe 
 a microscopic description to the physics at stake and the origin for,
at least, one further 
 equilibrium thermodynamic phase transition at some temperature below the paramagnetic transition at $T_c$.
Finally, and generally speaking, one may ask whether a discussion of ObD at $T=0^+$ is of pragmatic usefulness 
given that most experiments for which ObD may pertain typically proceed by cooling a material 
from a paramagnetic disordered phase to an ordered phase and not going below
a certain baseline nonzero temperature constrained by the experimental set-up.
This broader context provides the motivation for our work which aims to go beyond the sole consideration of a phenomenological
Ginzburg-Landau theory that contains all terms, relevant or not (in the renormalization-group sense),
allowed by symmetry and to expose how the different competing microscopic interactions participate to the degeneracy-lifting near the transition at $T_c$.

%Here, by understanding, we have in mind to expose, in a first instance, how the various couplings influence via their symmetry
%properties the topology of the phases selected near $T_c$ and their relative influence order of magnitude wise.

In the temperature regime near a phase transition, the harmonic approximation 
describing low-energy excitations usually  employed in theoretical discussions of ObD~\cite{shender_ObD,Henley_ObD,Shender_Peter,Yildirim_ObD}
at $T=0^+$ is not physically justified since large fluctuations typically accompany the phase transition to the paramagnetic state. 
From a fundamental viewpoint, it is thus highly desirable to study the role of fluctuations beyond s-MFT using a different method that can be applied at temperatures close to the critical region. 
A possible route to tackle this problem was paved by Thouless, Anderson and Palmer (TAP) in their study of the effect of fluctuations
 in spin glasses.~\cite{thouless1977solution} For the case of the Ising spin glass model, Plefka showed that the TAP correction 
can be systematically derived using a perturbative expansion;\cite{Plefka} an approach that was later generalized by Georges and Yedidia who calculated higher order terms in the Plefka's perturbative expansion.\cite{Yadidah} This approach, which  we refer to as extended TAP (E-TAP),  consists of a high-temperature expansion in $\beta\equiv 1/k_{\mathrm{B}}T$ about the s-MFT solution. It captures corrections to the s-MFT free-energy by including fluctuations  in the form of on-site linear and nonlinear susceptibilities.  A second approach allowing one to go beyond s-MFT is the cluster mean-field theory (c-MFT).\cite{Paramekanti,Yamamoto_CMFT}
 This approach treats short-range fluctuations within a finite cluster exactly while taking into account
 the inter-cluster interactions in a mean-field fashion. 

In the present work, we focus on the problem of ObD selection near criticality in a spin model on the pyrochlore lattice with predominant  
$xy$ interactions using the E-TAP method. 
We have recently used this method to investigate the problem of partial multiple-$\bm k$ order 
in pyrochlore magnets.~\cite{GTO-us} 
In this paper, we also employ c-MFT. We concentrate on the $\Gamma_5$ manifold for the $T \lesssim T^{\mathrm{MF}}_c$ regime since $\Gamma_5$ 
is not only an interesting theoretical playground according to recent investigations, 
\cite{pawel1,McClarty_Er2Ti2O7,Zhitomirsky:2012fk,Savary:2012uq,Wong,Ludo_edge,Petit_ObD,Zhitomirsky_danger} it is also of potential relevance to real pyrochlore
materials, such as Er$_2$Ti$_2$O$_7$, proposed to display an ObD 
mechanism.~\cite{Zhitomirsky:2012fk,Savary:2012uq, Champion_Er2Ti2O7,Ross_gap} 
In addition, the interesting case of distinct ObD selection at $T=0^+$ and $T_c$, 
reported in a pyrochlore system with anisotropic spin-spin coupling,~\cite{Chern:2010vn} 
further motivates us in investigating the role of the various spin interactions in the
state selection at $T\lesssim T_c$ in a general model of interacting spins on the  pyrochlore lattice.
% and pertinent to real materials.

The rest of the paper is organized as follows. In Section \ref{sec:mod}, we present the  bilinear nearest-neighbor spin model 
defined by the Hamiltonian ${\cal H}$ of Eq.~(\ref{h})  on the pyrochlore lattice and discuss its symmetries. 
Focusing on the $\Gamma_5$ manifold,
  we present a Ginzburg-Landau (GL) symmetry analysis to specify the general form of the lowest-order anisotropic terms allowed in the GL free-energy (${\cal F}_{\mathrm{GL}}$) that can lift the accidental $U(1)$ degeneracy found in s-MFT. 
We show that the fluctuation correction terms to the s-MFT free-energy select either the $\psi_2$ or the $\psi_3$ long-range ordered state of the $\Gamma_5$ manifold (see Fig. \ref{psi23} and Appendix \ref{apPsi23} for definition) 
as well as induce a weak moment $m_z$ along the local $[111]$ direction for the $\psi_2$ state. 
In Section \ref{sec:etap},  we present the E-TAP method and determine the phase boundary
at $T_c$ between the $\psi_2$ and $\psi_3$ states in the space of anisotropic spin-spin coupling constants. 
We investigate in Section \ref{sec:cmft}  how 
short-ranged fluctuations acting on the length-scale of a single-tetrahedron 
can lead to ObD when incorporated in the self-consistent scheme of c-MFT. The c-MFT method also allows us to explore 
 semi-quantitatively the role of nonzero effective Ising exchange ($\jzz$) on the selection at $T_c$. 
In Section \ref{multiple_trans}, we briefly discuss the possibility of multiple phase transitions in $xy$ pyrochlore magnets
 as temperature, for example, is varied.
Finally, we close with a discussion in Section \ref{sec:disc}. 
A number of appendices are provided to assist the reader with the technical details of the E-TAP calculations.
In Appendix \ref{apPsi23}, we detail the spin configurations of  the $\psi_2$ and $\psi_3$ states.
 In Appendix \ref{ap:symm}, we analyze the symmetry properties of the $\psi_2$ and 
the $\psi_3$ states to show that a fluctuation-induced local $m_z$ moment is only compatible with a $\psi_2$ long-range ordered state. 
We provide in Appendix \ref{apA} the  details of the E-TAP calculations and  the diagrammatic approach employed.

%==========================================================================================
%==========================================================================================

\section{Model}
\label{sec:mod}

We consider the following Hamiltonian on the pyrochlore lattice: \cite{Savary:2012uq,PhysRevX.1.021002}
%\addtocounter{footnote}{-1}\addtocounter{Hfootnote}{-1} \footnotemark
\begin{subequations}
\label{eq:hami}
\begin{eqnarray}
\mathcal{H}&=&\mathcal{H}_0+\mathcal{H}_1 \label{h}\\
\mathcal{H}_0&=&\sum_{\av{ij}}\jzz S_i^z S_j^z-\jp\pa{ S_i^+ S_j^-+ S_i^- S_j^+}\label{h0}\\
\mathcal{H}_1&=&\sum_{\av{ij}}\jpp\pa{S_i^+ S_j^+\gamma_{ij}+S_i^- S_j^-\gamma_{ij}^\ast}\notag
\\&+&\jzp\{S_i^z\pa{\zeta_{ij}S_j^++\zeta_{ij}^\ast S_j^-}+i\leftrightarrow j\}\label{h1}
\end{eqnarray}
\end{subequations}
where $S_i^{\pm} \equiv S_i^x \pm iS_i^y$ and $S_i^\mu$, with $\mu=z,+,-$, is defined in the local $[111]$ 
coordinate frame\cite{PhysRevX.1.021002} attached to each of the four pyrochlore sublattices (see Fig. \ref{psi23}). $\bm{S}_i$ can be treated classically as a 3-component vector or quantum mechanically as an operator such as a ${\bm S}_i=1/2$ pseudospin. In the context of magnetic rare-earth pyrochlores,
 $\bm S_i$ would represent either the total angular momentum $\bm J$
within a simplified model of ${\bm J}-{\bm J}$ coupling,~\cite{Hamid_TTO,Hamid_Paul_Mcihel,Petit_ObD,jordan_YbTO} 
or a pseudospin $1/2$ describing the single-ion 
ground state doublet.~\cite{Zhitomirsky:2012fk,Savary:2012uq,PhysRevX.1.021002,Gardner_RMP,Hamid_TTO,Hamid_Paul_Mcihel,jordan_YbTO,PhysRevLett.108.037202}

% depending on the energy spectrum  of the crystal electric field in the specific 
%material of interest.\cite{Zhitomirsky:2012fk,Savary:2012uq,PhysRevX.1.021002,Gardner_RMP,Hamid_TTO,Hamid_Paul_Mcihel,jordan_YbTO} 
%In the rest of this section, we treat $\bm S_i$ classicaly where $|\bm S_i|=1/2$ for all $i$. 

As we are foremost interested in the selection of classical ordered phases at $0 \ll T \lesssim T_c$, 
we shall treat $\bm S_i$ generally classically with $|\bm S_i|=1/2$ for all $i$.
 However, in Section \ref{sec:cmft}, where we use the c-MFT method, 
we consider $\bm S_i=1/2$ quantum mechanically, mostly for computational ease. 
In Eq. (\ref{eq:hami}), $\jzz,\jp,\jpp$ and $\jzp$ are the four 
symmetry-allowed independent nearest-neighbor exchange parameters, 
while $\zeta_{ij}=-\gamma_{ij}^\ast$ are bond-dependent phases on a single tetrahedron defined in
 Refs.~[\onlinecite{Savary:2012uq,PhysRevLett.108.037202,PhysRevX.1.021002}].
 {\footnote{It has recently been pointed out
by Y.-P. Huang, G. Chen and M. Hermele
[Phys. Rev. Lett. {\bf 112}, 167203 (2014)] that there are cases
(for example Nd$^{3+}$ in Nd$_2$Zr$_2$O$_7$ and Dy$^{3+}$ in Dy$_2$Ti$_2$O$_7$) where the 
single-ion ground state doublet is a so-called ``dipolar-octupolar''
 doublet in which $\gamma_{ij}$ is \emph{independent} of the $<i,j>$  bond. }}

It is important for the discussion that follows to split ${\cal H}$ into the two terms ${\cal H}_0$ and ${\cal H}_1$ in 
Eq.~(\ref{h})  and consider the symmetry properties of each term.
$\mathcal{H}_0$ has a $U(1)$ symmetry: it is invariant under a rotation of $\bm S_i$ by an arbitrary global angle 
 about the local $\langle 111 \rangle$ axes. 
On the other hand, the $\mathcal{H}_1$ term in Eq.~(\ref{h}) is only invariant under 
rotations of $\bm S_i$ by $\frac{2\pi}{3}$, reducing the symmetry of ${\cal H}$ to $Z_6$ ($C_3 [111] \times Z_2$).
%reduces this $U(1)$ symmetry to a $Z_6$ symmetry ($C_3 [111] \times Z_2$). 
%In other words, instead of rotation by an arbitrary angle, only 
%rotations of $\bm S_i$ by $\frac{2\pi}{3}$ around the local $\langle 111\rangle$ axes leave ${\cal{H}}_1$ invariant.  

%In the s-MFT treatment of a system with the Hamiltonian of Eq. (\ref{eq:hami}) \cite{Savary:2012uq}
%in a certain region of the parameter space centered around a dominant $\jp$ ($\Gamma_5$ region),
% the system orders in a state with a $\bm k=0$ 
%ordering wave vector with all the magnetic moments in the system lying  
%perpendicular to the local $[111]$ 
%direction and pointing in the same direction in the local coordinate system.
%the $xy$ plane of their local $[111]$ coordinate system 
%and pointing in the same local direction within that plane.

We consider a s-MFT treatment of the Hamiltonian of Eq.~(\ref{eq:hami}) which 
orders in a state with a ${\bm k} = 0$ 
ordering wave vector in a certain region of the parameter space 
centered around a dominant $\jp$ ($\Gamma_5$ region).
In this state, all the
magnetic moments in the system lie predominantly perpendicular to the
local $[111]$ direction and point in the same direction in the local coordinate system (see Fig.~\ref{psi23}).
These spin configurations define the $\Gamma_5$ manifold. \cite{Poole:2007ys} 
As a result, $\bm m_i\equiv\langle \bm S_i\rangle$, the on-site magnetization,{\footnote{ $<\cdots>$ 
represents a thermal Boltzmann average}} and $\phi_i\equiv\tan^{-1}(m_i^y/m_i^x)$, 
the azimuthal angle expressed in the local $[111]$ coordinate system, 
are independent of the lattice site index $i$. We henceforth drop the  index $i$ 
of $\phi_i$ and $\bm m_i$ for these ${\bm k}=0$ spin configurations.

As first noted in Ref.~[\onlinecite{Savary:2012uq}], in the $\Gamma_5$ region, the s-MFT free-energy 
is \emph{independent} of $\phi$  and the system displays an accidental $U(1)$ symmetry within such a s-MFT description.
 However, since ${\mathcal{H}}$ only has a global $Z_6$ symmetry, we expect that in a treatment 
of the problem that goes beyond s-MFT, this ``artificial'' $U(1)$ symmetry 
to be reduced to a $Z_6$ symmetry in the paramagnetic phase which gets spontaneously broken in the ordered phase. 
In this context, we note that a high-temperature series expansion of the quantum model (\ref{h}),
with values $\{J_\pm,J_{\pm\pm},J_{z\pm},J_{zz}\}$ appropriate for Er$_2$Ti$_2$O$_7$,~\cite{Savary:2012uq}
shows explicitly that such a $Z_6$ anisotropy develops \emph{in the paramagnetic phase} 
upon approaching $T_c$ from above. \cite{Oitma1}
Also, a recent Monte Carlo study has investigated the critical behavior of a classical version of this model. \cite{Zhitomirsky_danger}
We now present a Ginzburg-Landau (GL) symmetry analysis that allows one to anticipate 
the form of the lowest order fluctuation corrections to s-MFT that lift the $U(1)$ degeneracy.

%%%%%%%%%%%%%%%%%%%%%%%%%%%%%%%%%%%%%%%%%%%%%%%%%%%%%%%%%%%%%%%%%%%%%%%%%%%%%%%%%%%%%%%%%%%%%%%%%%%%%%%%

\subsection{Ginzburg-Landau (GL) symmetry analysis}
\label{GLanalysis}

We start by defining the complex variable
 \begin{equation}
 \label{op3}
 m_{xy}\equiv\sqrt{m_x^2+m_y^2}\exp(i\phi),
 \end{equation}
 which corresponds to the magnitude and direction of the on-site magnetization $\bm m = \langle \bm S_ i\rangle$  in the local $xy$ plane. 
 
  For simplicity, we first consider a strictly local-$xy$ state and assume $m_z=0$. 
The allowed terms in the Ginzburg-Landau (GL) free-energy can only be of even powers
 in $m_{xy}$ due to  time-reversal symmetry, $\tau$.  On the other hand,
the rotation by $2\pi/3$ about the cubic $\langle111\rangle$ axes, or $C_3$ symmetry,
forbids the existence of terms of the form $m_{xy}^n+(m^{\ast}_{xy})^n$ unless $n$ is a multiple of three. 
The resulting terms have the ability to lift the accidental $U(1)$ degeneracy of the $\Gamma_5$ manifold since 
they introduce a dependence of the GL free-energy on the azimuthal angle $\phi$, 
the orientation of $\bm m$ in the local $xy$ plane (see Eq. (\ref{op3})).
 Considering the effect of $C_3$ and $\tau$ together, 
assuming strictly $xy$ order ($m_z=0$), 
the lowest order term that breaks the $U(1)$ symmetry  must therefore have $n=6$ and be of the form
 \begin{equation}
 \label{GL1}
\eta_6(T) [ m_{xy}^6+(m^{\ast}_{xy})^6 ], 
 \end{equation}
where the anisotropy strength $\eta_6$ depends on temperature, $T$.
% as well as magnitude and direction of the applied magnetic field ${\bm h}$.
 We note that such a sixth order term is dangerously irrelevant for the 3-dimensional $xy$ universality 
class.{~\footnote{It is dangerously irrelevant since at temperatures $T < T_c$ 
this becomes relevant above a length scale $\lambda$, which diverges as a power of 
the correlation length.~\cite{Zhitomirsky_danger,Bal-San-3Dxy}}}
  This means that while it does not affect the critical properties of the system,\cite{Bal-San-3Dxy,Zhitomirsky_danger} 
 this term plays a crucial role in the selection of a specific long-range ordered state 
by lifting the $U(1)$ s-MFT degeneracy at $T \lesssim T_c$.
Higher order terms $f_{6n} \sim {\vert m_{xy}}\vert ^{6n} $ ($n>1$) 
are also generated by fluctuations, but are even
more irrelevant than $f_6$ at $T_c$.
As an example, we discuss in Section \ref{multiple_trans} the effects of competing $f_{6}$ and $f_{12}$ terms
at $T<T_c$ as the amplitude $\vert m_{xy}\vert$ grows upon cooling below $T_c$.

If we now include the $z$ component of the order parameter, i.e. $m_z\neq 0$, a $U(1)$ 
symmetry-breaking term arises at fourth order in the components of $\bm m$ in the GL free-energy.
Such a term was only recently noted in a numerical study~\cite{Petit_ObD} but whose 
microscopic and symmetry origins was not discussed.
%which has not been previously considered.~\cite{pawel1,Zhitomirsky:2012fk,Savary:2012uq,Wong}
Again, based on the combined effect of the $C_3$ and $\tau$ symmetry operations, 
the degeneracy-lifting fourth order term in the GL free-energy has the form 
 \begin{equation}
 \label{GL2}
 \omega(T) m_z [ m_{xy}^3+(m^{\ast}_{xy})^3] ,
 \end{equation}
where, here again, the anisotropic coupling $\omega$ depends on $T$.
Together, Eqs. (\ref{GL1}, \ref{GL2}) identify the form of the two lowest order terms 
in the GL free-energy capable of lifting the degeneracy 
within the $\Gamma_5$ manifold.  

It turns out that the sixth order term of Eq.~(\ref{GL1}) and the  fourth order term of Eq.~(\ref{GL2})
 have the same net effect in lifting the degeneracy beyond s-MFT as can be shown by combining them 
into a single term in the GL free-energy, ${\cal F}_{\mathrm {GL}}$. 
Consider first the  quadratic terms in ${\cal F}_{\mathrm {GL}}$ of the form $r_0(m_x^2+m_y^2)+r_1m_z^2$.
Here $r_0$ and $r_1$ are chosen to be different
%to incorporate the physics arising from the microscopic anisotropic exchange 
%in the Hamiltonian of  Eq. (\ref{eq:hami}) and 
to emphasize the distinct criticality for 
the $xy$ and $z$ components of the order parameter $\bm m$. 
Next, taking into account the terms of Eqs. (\ref{GL1}, \ref{GL2}), we have:
\begin{eqnarray}
\label{GLfin1}
{\cal F}_{\mathrm {GL}}&=&r_0(m_x^2+m_y^2) + r_1m_z ^2 + \omega m_z [m_{xy}^3+{(m_{xy}^{\ast})}^3] 	\nonumber\\
&& + \eta_6 [ (m_{xy}^6+(m^{\ast}_{xy})^6 ] + {\cal F}_{\mathrm {GL}}^{(4)}(m_{xy},m_z)\nonumber\\
&& + {\cal F}_{\mathrm {GL}}^{(6)}(m_{xy},m_z)+ \cdots .
\end{eqnarray}
This can be rewritten as
\begin{eqnarray}
\label{GLfin2}
{\cal F}_{\mathrm {GL}}&=&r_0(m_x^2+m_y^2) + r_1(m_z + \frac{\omega|m_{xy}|^3\cos(3\phi)}{r_1})^2\nonumber\\
&&- \frac{\omega^2|m_{xy}|^6}{2r_1}+( 2\eta_6 - \frac{\omega^2}{2r_1})|m_{xy}|^6\cos(6\phi)\nonumber \\
&& + {\cal F}_{\mathrm {GL}}^{(4)}(m_{xy},m_z)+{\cal F}_{\mathrm {GL}}^{(6)}(m_{xy},m_z)+ \cdots ,
\end{eqnarray}
where we used Eq. (\ref{op3}) to go from  Eq.~(\ref{GLfin1}) to Eq.~(\ref{GLfin2}).
In Eqs.~(\ref{GLfin1},\ref{GLfin2}), ${\cal F}_{\mathrm {GL}}^{(4)}$ and ${\cal F}_{\mathrm {GL}}^{(6)}$ 
are fourth and sixth order terms
%that {\it do not} lift the $U(1)$ degeneracy
and $\cdots$  represents terms that are of higher order in the components of $\bm m$. 
We take $r_1> 0$ to enforce no criticality for the $m_z$ component of $\bm m$.
Upon minimizing Eq. (\ref{GLfin2}) with respect to $m_z$, we obtain:
\begin{equation}
\label{mz-GL}
 m_z=-\omega |m_{xy}|^3\cos(3\phi)/r_1.
 \end{equation}
Hence, after having minimized ${\cal F}_{\mathrm{GL}}$ 
with respect to $m_z$, we have
\begin{equation}
\label{GLfin}
\delta {\cal F}_{\mathrm {GL}}(\phi) \equiv 2 \bar \eta_6 |m_{xy}|^6\cos(6\phi),
\end{equation}
with
\begin{equation}
\label{eq:bareta6}
2\bar \eta_6 \equiv (2\eta_6 - \omega^2/2r_1),
\end{equation}
and where $\delta {\cal F}_{\mathrm {GL}}(\phi)$ is the anisotropic part of ${\cal F}_{\mathrm {GL}}$, 
 the $U(1)$ degeneracy-lifting term on which we focus.
The minimum of $\delta {\cal F}_{\mathrm {GL}}$  
is either $\cos(6\phi)=\pm 1$ depending on the sign of $\bar \eta_6$.
This happens for $\phi=n\pi/3$   or $\phi=(2n+1)\pi/6$, with $n=0,1, \ldots, 5$ with
these two sets of angles corresponding, respectively, to the $\psi_2$ and the $\psi_3$ states (see Fig. \ref{psi23}). 
So, when $\bar \eta_6 < 0$ ($\bar \eta _6 > 0$), the minimum free-energy state is $\psi_2$ ($\psi_3$). 
The boundary between  $\psi_2$ and $\psi_3$ is thus determined by the real roots of  $\bar \eta_6$.
A similar discussion is invoked in Ref.~[\onlinecite{Wong}] to describe 
the zero temperature  $\psi_2$-$\psi_3$ phase boundaries arising from quantum ObD.

We note that the fluctuation corrections to s-MFT induce a local $m_z$ moment to the minimum 
free-energy state which was found to be strictly ordered in the
 local $xy$ plane at the 
s-MFT level.~{\footnote{We note that, as we were concluding the present work, 
a very recent paper by Petit \emph{et al.}\protect{\cite{Petit_ObD}}
building on the model of \protect{Ref.~[\onlinecite{McClarty_Er2Ti2O7}]},
also reported a weak $m_z$ moment in a mean-field theory calculation considering anisotropic
${\bm J}-{\bm J}$ couplings and the crystal electric field of Er$_2$Ti$_2$O$_7$.
}
}
Since this moment is proportional to $\cos(3\phi)$, only the $\psi_2$ state can have a nonzero $m_z$. 
This result is also expected based solely on symmetry considerations for
the $\psi_2$ and $\psi_3$ states (see Appendix \ref{ap:symm}).
The induced moment is, however,  small just below $T_c$ since 
it is proportional to $|m_{xy}|^3$ and inversely proportional to $r_1$ 
which remains firmly positive away from spontaneous Ising criticality at $r_1=0$.

 At the phenomenological GL level, Eq. (\ref{GLfin}) is the final result demonstrating 
how the $U(1)$ degeneracy at the s-MFT level may be lifted by the anisotropic terms of Eqs.~(\ref{GL1},\ref{GL2}). 
In this work, however, we are rather interested in 
exposing how the coefficient of the symmetry-breaking terms in the free-energy of Eq. (\ref{GLfin2}), $\bar \eta_6$, 
can be determined \emph{at a microscopic level} when going beyond  a s-MFT description. 
Specifically, we wish to explore the leading dependence of $\bar \eta_6$ upon the
$J_{\pm\pm}$, $J_{z\pm}$ and $J_{zz}$ anisotropic exchange couplings near $T_c$.

%================================================================================================
%================================================================================================
\section{Methods and Results\label{section:method}}

In Section \ref{sec:etap}, we use the E-TAP method to compute
the phase boundary between the $\psi_2$ and $\psi_3$ states determined by the function 
$\bar \eta_6$ in Eq.~(\ref{GLfin}) for $T \lesssim T_c^{\mathrm{MF}}$.
 In this calculation,  we consider for simplicity $\jzz=0$ in Eq. (\ref{eq:hami}). {\footnote{Considering $J_{zz} \neq 0$ contributions to the
 $\cos(6\phi)$ coefficient would require calculating higher order terms in $\beta$ in the expansion of Eq. (\ref{Gibbs1}). Such a 
calculation is beyond the scope of the present work.}} In Section \ref{sec:cmft}, we conduct a complementary numerical study of fluctuation corrections to the s-MFT using cluster-MFT (c-MFT) which allows us to explore the effect of fluctuations at the level of one tetrahedron for
both $J_{zz}=0$ and $J_{zz} \neq 0$. Since we are essentially interested in the $\Gamma_5$ region where the dominant interaction is $\jp$ 
 in Eq. (\ref{eq:hami}), 
we express the rest of the couplings in units of $\jp$ 
and we henceforth denote the scaled (perturbative) interactions with lower case letters, i.e. 
$j_{\pm\pm}\equiv \jpp/\jp$  and $j_{z\pm}\equiv J_{z\pm}/\jp$.

 \subsection{Extended-TAP Method (E-TAP)}\label{sec:etap}
\subsubsection{ Method} \label{sec:modetap}

In a magnetic system with static magnetic moments, a given moment is subject to 
a local field due to its neighbors. In a s-MFT treatment, the moment affects
its own local field indirectly; this is an artifact of s-MFT.
The so-called Onsager reaction field introduces a term in the s-MFT free-energy
that cancels this unphysical effect to leading order. 
%This correction, first introduced by Thouless, Anderson and Palmer 
%(TAP)\cite{thouless1977solution}, 
%is particularly important in setting up a proper mean-field theory description 
%of spin glasses with infinite-range interactions.\cite{Fisher-Hertz}
%In 1977, Thouless, Anderson and Palmer (TAP)\cite{thouless1977solution} 
%proposed a method to study the role of fluctuations in spin glasses. 
%This method is equivalent to calculating the Onsager reaction field. \cite{naiveTAP} 
As was shown by  Thouless, Anderson and Palmer (TAP),~\cite{thouless1977solution} 
this reaction-field correction is particularly important in
setting up a proper mean-field theory description 
of spin glasses with infinite-range interactions.\cite{Fisher-Hertz}
Here, we present an extended version of the TAP method (E-TAP) first developed by Georges and Yedidia\cite{Yadidah} for Ising spin glasses
and which includes the lowest order fluctuation corrections originally calculated by TAP as well as those beyond.

To proceed, we must first modify the E-TAP procedure 
of Ref. [\onlinecite{Yadidah}] to study the case of 3-component classical spins with anisotropic exchange interactions. 
In the E-TAP method, nonzero on-site fluctuations i.e. 
$ \langle S^{\alpha}_{i} S^{\beta}_{i}\rangle \neq \langle S^{\alpha}_{i} \rangle \langle S^{\beta}_{i}\rangle$,\cite{Yadidah} 
are taken into account via a high-temperature expansion (small $\beta$) of a Gibbs free-energy:
 \begin{equation}
\label{Gibbs1}
G=-\frac{1}{\beta}\ln\pa{\text{Tr}[\exp\big(-\beta \mathcal{H}+ \sum_i \bm \lambda_i \cdot (\mathbfit S_i-\mathbfit m_i)\big)]}.
\end{equation}
Here, $\mathbfit m_i$ is the average magnetization at site $i$, $\mathbfit m_i\equiv\langle \mathbfit S_i \rangle$  and $\bm \lambda_i$ is a Lagrange multiplier which fixes $\mathbfit m_i$ to its mean-field value.
The high-temperature expansion introduces fluctuations about the s-MFT solution.
Defining $\beta G(\beta)\equiv \tilde{G}(\beta)$, the first two terms of the expansion in powers of $\beta$,
 $\tilde{G}(0)/\beta$ and $\tilde{G}^\prime(0)$, are the entropy and energy at the s-MFT level, respectively. 
The prime represents differentiation with respect to $\beta$. 
 The higher order terms in the $\beta$ expansion of the Gibbs free-energy correspond to fluctuation corrections to the s-MFT free-energy that generate the terms that lift  the $U(1)$ degeneracy. 
We aim to calculate the higher order terms (beyond $\beta^2$) in the $\beta$ expansion of Eq. (\ref{Gibbs1}) that contribute to the degeneracy lifting term $2\bar\eta_6 |m_{xy}|^6\cos(6\phi)$ of Eq. (\ref{GLfin})
 to lowest order in the coupling constants $\jspp$ and $\jszp$. {\footnote{We recall that $\jp$ is the dominant interaction and $\jspp$ and $\jszp$ are being treated perturbatively in Eq. (\ref{eq:hami}).}}  
In other words, we are considering the selection 
near mean-field criticality and perturbative (in $j_{z\pm}$ and $j_{\pm\pm}$)  vicinity of the $U(1)$ symmetric portion of the theory
($H_0$ in Eq. (\ref{h})). 
Since we are focusing on the expansion of $G$ in Eq. (\ref{Gibbs1}), we write its correction $\delta G$ beyond the s-MFT solution
 as suggested by Eq.(\ref{GLfin}):
 \begin{equation}
 \label{freefinal}
 \delta G =  2\bar \eta_6 (\jspp, \jszp)|m_{xy}|^6\cos(6\phi).
 \end{equation}
As discussed in Section \ref{GLanalysis}, $\bar \eta_6(\jspp, \jszp)=0$ determines
 the phase boundaries between the $\psi_2$ and the $\psi_3$ states in the space of coupling constants $\jspp$ and $\jszp$.
 We now proceed to calculate $\bar\eta_6(\jspp, \jszp)$ with the E-TAP method.

In the $n^\mathrm{th}$ order of the $\beta$ expansion, factors of the form $\jspp^k\jszp^l$ 
arise where $k$ and $l$ are positive integers and $k+l=n$. Each power of $\jspp$ and $\jszp$ 
contributes factors of $e^{\pm 2i\phi}$ and $e^{\pm i\phi}$, respectively (see Eq. (\ref{h1})), to the corresponding term in the $\beta$ expansion. 
By a simple power counting of these factors, one can pinpoint the terms that contribute to $\cos(6\phi)$ 
and $\cos(3\phi)$ that are necessary to compute due to their 
 contribution to $\omega$ and $\eta_6$ and, consequently, to $\bar\eta_6$ (see. Eq.~(\ref{eq:bareta6})),
 and which arise at different orders  in $\beta$ in the E-TAP calculation.
 It is straightforward 
arithmetic to find what combinations of $k$ and $l$ in $\jspp^k\jszp^l$ generate $\cos(6\phi)$ and $\cos(3\phi)$ terms.
We then calculate $\omega$ and $\eta_6$ in Eqs. (\ref{GLfin1}, \ref{GLfin2}) in terms of the microscopic couplings $\jspp$ and $\jszp$
using the E-TAP method (see Appendix \ref{apA}). We obtain $\omega(\jspp, \jszp)$, 
$\eta_6(\jspp, \jszp)$,  and thus $\bar \eta_6(\jspp, \jszp)$, all to  lowest nontrivial order in $\jspp$ and $\jszp$.
These read 
 as:{\footnote{We note that E-TAP corrections of the coefficient $r_1$ in
Eq. (\ref{GLfin1}) does not contribute to $\bar \eta_6(\jspp, \jszp)$ when it 
is calculated to the lowest order in the coupling constants $\jspp$ and $\jszp$ and we therefore use its s-MFT $r_1=2/\beta$ value.}} 
 \begin{widetext}
\begin{equation}
\label{eq:g0}
\omega(\jspp,\jszp) = a_0\beta\jspp\jszp+\beta^2(a_1\jspp^2\jszp+ a_2\jszp^3)+\beta^3(a_3\jspp^3\jszp+a_4\jspp\jszp^3) 
\end{equation}
and
\begin{equation}
\label{eq:g1}
\eta_6(\jspp,\jszp) = b_0\beta^2\jspp^3+b_1\beta^3\jspp^2\jszp^2+ b_2\beta^4\jspp\jszp^4+b_3\beta^5\jszp^6,
\end{equation}
and, consequently,
\begin{equation}
\label{eq:mainC}
2\bar\eta_6(\jspp,\jszp) = c_0\beta^2\jspp^3+c_1\beta^3\jspp^2\jszp^2+ c_2\beta^4\jspp\jszp^4+ \beta^5(c_3\jspp^4\jszp^2+c_4\jspp^2\jszp^4+c_5\jszp^6).
\end{equation}

\end{widetext}
In Eqs. (\ref{eq:g0}, \ref{eq:g1}, \ref{eq:mainC}), $a_i$ ($i=0,\cdots, 4$),
 $b_j$ ($j=0,\cdots, 3$) and $c_k$ ($k=0,\cdots,5$)
 are numerical coefficients determined by the explicit E-TAP calculation too lengthly to reproduce here, but whose
derivation is described  in Appendix \ref{apA}. 
Considering the highest power of coupling constants $\jspp$ and $\jszp$ in Eq. (\ref{eq:mainC}), one needs to compute terms up to sixth order in the high-temperature expansion of $G$ in Eq. (\ref{Gibbs1}).

\subsubsection{Results}

\begin{figure}
\centering
\includegraphics[scale=0.9]{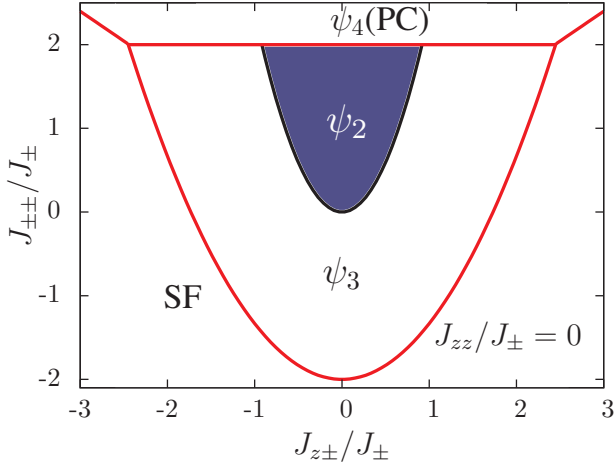}
\caption{(Color Online)
 The $\jszz=0$ phase diagram of the model at $T\lesssim T_c$.
 The $\Gamma_5$ region corresponds to the region enclosing the $\psi_2$ and $\psi_3$ states taken together. 
The $\Gamma_5$ region is circumscribed by an outer (red) parabola $\jspp=2\jszp^2/3-2$
 and a horizontal red line at $\jspp=2$ obtained from the s-MFT calculation.\cite{Wong} 
The phase boundary between the $\psi_2/\psi_3$ states obtained from the E-TAP calculations
(i.e. the roots of $\bar \eta_6=0$, see text) is represented by the black line. 
The phases outside of the boundaries of the $\Gamma_5$ region are: a splayed ferromagnet (SF)~\cite{PhysRevX.1.021002}
 canted from the $[100]$ cubic direction and the so-called $\psi_4$ or Palmer-Chalker (PC) state.
\cite{Wong,Poole:2007ys,PC1}
Both the SF and the PC phases have ${\bm k=0}$ ordering wave vector.
\label{fig:Gamm5}
}
\end{figure}

In order to obtain the numerical value of the $a_i$, $b_j$ and $c_k$ coefficients in Eq. (\ref{eq:mainC}), 
we employ a diagrammatic technique to represent the terms in the $\beta$ expansion and 
which constitutes the computational core of the E-TAP method. These diagrams are composed of vertices and bonds. \cite{oitmaa} 
The vertices correspond to lattice sites covered by a diagram and the bond represents the interaction between the vertices that it connects. The details necessary to carry out the calculations using diagrams are presented in Appendix \ref{sec:diagrams}. 
As mentioned earlier, for computational simplicity, we only consider the case of $j_{zz}=0$.
% since the coupling between the pseudospins within $\Gamma_5$ is mainly $xy$. 
Recalling Eq.~(\ref{eq:bareta6}), and noting that to the lowest order we can use the s-MFT value of $r_1$, $r_1=2/\beta$, we find:
 \begin{widetext}
 \begin{eqnarray}
 \label{cubic-res2}
2\bar\eta_6(\jspp,\jszp)&=& 10^{-5}\times[-6.50\beta^2 \jspp^3+57.3\beta^3\jspp^2\jszp^2+26.5\beta^4\jspp\jszp^4+27.8\beta^5\jszp^6\nonumber \\
&&-7.95\beta^5\jspp^4\jszp^2-4.60\beta^4\jspp^3\jszp^2+66.5\beta^5\jspp^2\jszp^4] .
 \end{eqnarray}
 \end{widetext}
In deriving Eq.~(\ref{cubic-res2}), 
we used $\omega$ and $\eta_6$ in Eqs. (\ref{eq:g0}, \ref{eq:g1}), respectively, 
with the numerical values of $a_i$ and $b_j$ coefficients given in Tables \ref{tab:etap-g0-a} and \ref{tab:etap-g1-b}.
\begin{table}
\begin{center}
    \begin{tabular}{ | l | l | l | l | l |}
    \hline
    $a_0$ & $a_1$  & $a_2$  & $a_3$ & $a_4$ \\ \hline
    -0.593& -1.13 &-0.451& -0.938 & 7.21 \\ \hline
    \end{tabular}
   \caption{Values of $a_i$ in $\omega(\jspp, \jszp)$ multiplied by $10^2$.}
  \label{tab:etap-g0-a}
\end{center}
 \end{table}  
 \begin{table}
\begin{center}
    \begin{tabular}{ | l | l | l | l | }
    \hline
    $b_0$ & $b_1$  & $b_2$  & $b_3$  \\ \hline
    -0.322 & 2.92 &1.42 & 1.42  \\ \hline
    \end{tabular}
   \caption{Values of $b_j$ in $\eta_6(\jspp, \jszp)$ multiplied by $10^4$.}
  \label{tab:etap-g1-b}
\end{center}
 \end{table}  
Since $\bar \eta_6(\jspp, \jszp)=0$ determines the boundaries between the $\psi_2$ and $\psi_3$ states, we 
need to compute the roots of Eq.~(\ref{cubic-res2}). 
In order to be consistent with the above E-TAP derivation perturbative in $j_{z\pm}$ and $j_{\pm\pm}$, 
we ought to only consider the lowest order term in the root of $\eta_6$ given by Eq.~(\ref{cubic-res2}).
We find that Eq. (\ref{cubic-res2}) has one real root which to, 
the lowest order in the coupling constants, reads:
\begin{equation}
\label{root}
\jspp \approx p_0\jszp^2 + \cdots
\end{equation}
where the $\cdots$  represent higher order terms in $\jszp$ and $p_0 \approx 9.37\beta$
which determines the phase boundary between the $\psi_2$ and the $\psi_3$ states for $\jszz=0$, $j_{z\pm} \ll 1$ 
and $\beta \gtrsim \beta_c$.  
%The $\psi_2/\psi_3$ boundary is shown Fig. \ref{fig:Gamm5} for $\beta=0.253 \gtrsim \beta_c^{\mathrm{MF}}=1/4$. 
The corresponding $\psi_2$-$\psi_3$ boundary for $\beta=0.253 \gtrsim \beta_c^{\mathrm{MF}}=1/4$,
providing an estimate of the boundary in the limit $T\rightarrow T_c^{-}$, is shown Fig.~\ref{fig:Gamm5}
The ``outer''  s-MFT boundaries of the overall $\Gamma_5$ region for $\jszz=0$ is also shown in this figure.
Outside the $\Gamma_5$ boundary, the system orders in a Palmer-Chalker (PC) state \cite{Wong,Poole:2007ys,PC1}
 or a splayed ferromagnet (SF).~\cite{PhysRevX.1.021002}

It was found in Ref.~[\onlinecite{Wong}] 
that, at zero-temperature, quantum fluctuations yield three distinct phase boundaries for $\jszz = 0$, 
a result confirmed in Ref.~[\onlinecite{Ludo_edge}],
even in the regime $j_{z\pm}\ll 1$ and $j_{\pm\pm} \ll 1$, for which the E-TAP results 
for the ordered state selection at $T\lesssim T_c$ presented here apply.
 The combination of the E-TAP results with those from Ref. [\onlinecite{Wong}] 
suggests, because of the different $\psi_2$/$\psi_3$ boundaries at $T=T_c^-$ and $T=0$,
the possibility of multiple transitions between $\psi_2$ and $\psi_3$ as the temperature is decreased well below $T_c$ as
was found for the model of  Ref.~[\onlinecite{Chern:2010vn}].
 Such a multiple-transition scenario constitutes an exotic variant of the more conventional ObD phenomenon
 since fluctuations select distinct long-range ordered states in different temperature regimes ($T \lesssim T_c$ and $T=0^+$).
That being said, one should be reminded 
that since the E-TAP phase diagram of Fig. \ref{fig:Gamm5} was constructed on the basis of a {\it lowest order}  
fluctuation correction to s-MFT described by Eqs.~(\ref{eq:g0},\ref{eq:g1},\ref{eq:mainC})
and for a classical version of ${\cal H}$ in Eq.~(\ref{eq:hami}), 
this discussion of multiple transitions is therefore only qualitative within the present E-TAP calculation.
However, we expand further on this topic in Section \ref{multiple_trans} within a Ginzburg-Landau theory framework.
In the low-temperature regime, $T\ll T_c$, 
a study that incorporates high-order magnon-magnon interaction at nonzero temperature starting
 from the results of Ref. [\onlinecite{Wong}] would be of interest to explore this phenomenology further. 
This is, however, beyond the scope of the present work.

%===================================================================================================
%===================================================================================================

\subsection{Cluster mean-field theory (c-MFT)}
\label{sec:cmft}

It is interesting to ask to what extent a calculation, any calculation, that goes beyond s-MFT may reveal an ordered state selection
 at $T\lesssim T_c$. For this reason, we study in this section the effect of fluctuations numerically using c-MFT. 
This method incorporates fluctuations at the level of one tetrahedron by exactly diagonalizing the Hamiltonian corresponding to the tetrahedron and treating the interactions between different tetrahedra at the s-MFT level. The c-MFT approach does not involve a perturbative scheme in powers of $\beta$ compared with the E-TAP method where fluctuations are included through an expansion of the Gibbs free-energy in powers of $\beta$ (see Eq. (\ref{Gibbs1})). In c-MFT, fluctuations are limited to one tetrahedron while, in the E-TAP method, fluctuations beyond one tetrahedron are included up to the range spanned by the diagrams generating the sixth order terms in the high-temperature expansion of $G$ (see Appendix \ref{sec:diagrams} for details). One practical advantage of the c-MFT method compared to the E-TAP approach is that c-MFT allows us to easily investigate the effect of nonzero $J_{zz}$ on the selection of the $\psi_2$ versus the $\psi_3$ state.  In what follows we first provide the details of the  c-MFT method 
 and then present in Fig.~\ref{fig:cMFT} the results of the calculations for various $\jzz$ values.

\subsubsection{Method} 

% In order to account for the minimal magnetic properties and symmetries of the pyrochlore lattice while 
% incorporating fluctuation corrections to s-MFT, we employ a cluster mean-field theory (c-MFT) method. 

\begin{figure}
\centering
\includegraphics[scale=0.30]{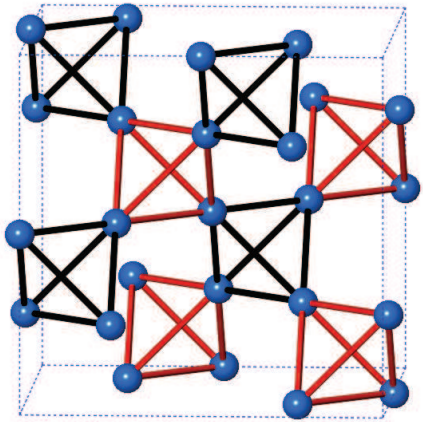}
\caption{
The pyrochlore lattice, with spins located at the vertices of corner-sharing tetrahedra. There are two orientations of tetrahedra indicated by the black and red colored tetrahedra with the interaction paths represented by the colored lines. The centers of the tetrahedra form a diamond lattice (with sublattice A and B corresponding respectively to,
say, the black and red tetrahedra).
 In c-MFT, the interaction in-between the black tetrahedra is treated at the  mean-field theory level.
}
\label{pyro}
\end{figure}

The c-MFT that we  use here may be viewed as a general, non-perturbative and rather system-independent
 way to obtain (numerical) results beyond s-MFT. While the approach is not specific to the type of spins
(classical or quantum)  considered, we restrict ourselves to quantum spins $1/2$ for a tetrahedron cluster.
% Hence our results are based on a numerical exact diagonalization (ED) of a tetrahedron cluster 
% where $\Sl_i^{\mu}$ is taken as a quantum mechanical 
%operators with ${\bm S_i}=1/2$ and comprise short-range fluctuations within one tetrahedron. 
A pragmatic reason for using quantum spins here is to avoid the complicated eight integrals over the solid angle that correspond to the classical trace for classical spins taken as vectors of fixed length $|\bm{S}_i|=1/2$ and orientation defined by an azimuthal and a polar angle. To illustrate the method, 
 we write the model Hamiltonian \eqref{eq:hami} in the compact form:
\eq{
\H=\sum_{\av{ij},\mu,\nu}J_{ij}^{\mu\nu}\Sl_i^\mu \Sl_j^\nu,\label{sav}
}
with $\mu,\nu=z,+,-$ and:
\eq{	
J_{ij}^{\mu\nu}:=\left(\begin{array}{ccc}J_{zz} & J_{z\pm} \zeta_{ij} & J_{z\pm}\zeta_{ij}^* \\J_{z\pm} \zeta_{ij} & J_{\pm\pm}\gamma_{ij} & -J_{\pm} \\J_{z\pm}\zeta_{ij}^* & -J_{\pm} & J_{\pm\pm}\gamma_{ij}^*\end{array}\right).
}
 For $N$ sites on the pyrochlore lattice, there are $N/4$ "up" and $N/4$ "down" corner-sharing tetrahedra (see Fig. \ref{pyro}). The diamond lattice, dual to the pyrochlore lattice, offers a simple representation of the tetrahedra 
that  tessellate  the lattice as elementary units.
We note that the diamond lattice is bipartite (it is composed of 2 FCC sublattices, say A and B). As we consider ordered phases with a $\mathbfit{k}=0$ ordering wave vector, we assume in the c-MFT method that all tetrahedra on sublattice A are equivalent and interact with each other only through the mean fields. 
We re-express model \eqref{sav} in a more suggestive form as a sum over sublattice A (labelled with $I$ or $I'$) and take $i,j=1,2,3,4$ as the sublattice indices:
\eq{
\H=\sum_{I\in A}\sum_{\av{ij},\mu,\nu}J_{ij}^{\mu\nu}\Sl_{i,I}^\mu \Sl_{j,I}^\nu+\sum_{\av{I,I'}\in A}\sum_{\av{ij},\mu,\nu}J_{ij}^{\mu\nu}\Sl_{i,I}^\mu \Sl_{j,I'}^\nu\label{mfdec}
}

We then proceed to apply a s-MFT approximation {\it only} on the second term of the Eq.\eqref{mfdec}: we neglect any fluctuations \emph{between} the A tetrahedra while taking full account of fluctuations \emph{within} the A tetrahedra. The self-consistently determined mean fields $\av{S_{i,I}^\mu}\equiv m_{i,I}$ are introduced to decouple the inter-cluster bonds. Again, because we focus on the $\Gamma_5$ manifold with $\mathbfit{k}=0$ ordering wave vector, we have translational invariance of sublattice A which implies $m_{i,I'}^\nu=m_{i,I}^\nu$. By performing this approximation, Eq. \eqref{mfdec} reduces to a sum over the A tetrahedra coupled together by the mean-fields:
\eq{
\Hm&=\sum_{I\in A}\sum_{i,j,\mu,\nu}J_{ij}^{\mu\nu}\pa{\frac{1}{2}\Sl_{i,I}^\mu \Sl_{j,I}^\nu+\Sl_{i,I}^\mu m_{j,I}^\nu-\frac{1}{2}m_{i,I}^\mu m_{j,I}^\nu}.\label{hmf}
}
Any thermodynamic average is readily computed from $\Hm$. In particular, 
it is straightforward to show that the physical solution $\ac{m_i^\mu}$ that 
minimizes the free-energy corresponds to $m_i^\mu=\av{\Sl_i^\mu}$, where the thermodynamic average is taken with respect to the cluster mean-field Hamiltonian $\H_{\mathrm{MF}}$. While we focus on the $|\bm S_i|=1/2$ case, we note that c-MFT could be used for $|\bm S_i|>1/2$ and extended to clusters with more than 4 spins. The method (using quantum spins) is only constrained by the computational limitations of the required exact full diagonalization over the cluster considered.\cite{2010AIPC.1297..135S}

\subsubsection{Results}

The critical temperature $T_c$ is a function of  the $J_{\pm}$, $J_{z\pm}$ and $J_{\pm\pm}$ exchange parameters.
 For every point in the phase diagrams (i.e. for a given set of  exchange couplings $j_{z\pm}$ and $j_{\pm\pm}$),
we determine $T_c$ by identifying the temperature at which
there is a minimum of the free-energy and the development of a numerically non-zero on-site magnetization ${\bm m}$.
For definitiveness, we then take a temperature slightly below $T_c$, $T=0.9 T_c$,
 for each ($j_{z\pm},j_{\pm\pm}$) and determine which state
($\psi_2$ or $\psi_3$) is selected. We record this state selection over a grid of points that span the $\Gamma_5$ manifold.

\begin{figure}[h!]
\centering
\includegraphics[scale=0.8]{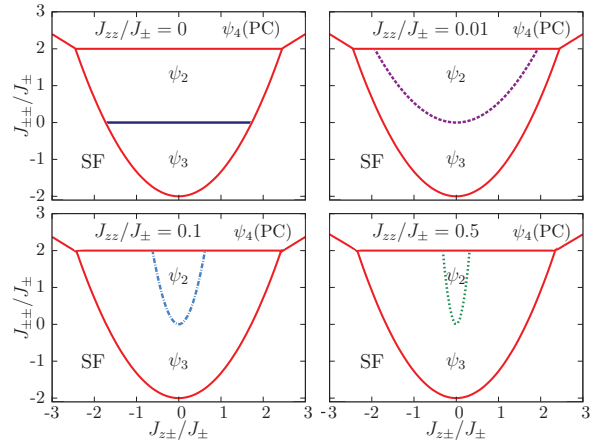}
\caption{
(Color Online) The c-MFT phase diagram at $T\lesssim T_c$ obtained for positive 
$\jszz=0.5,0.1,0.01,0$. The region circumscribed by the red line is the $\psi_2/\psi_3$ $\Gamma_5$ region. 
For a given phase boundary (fixed $\jszz$) the area above (below) that phase boundary line corresponds to the 
$\psi_2$ ($\psi_3$) phase. The phase boundaries for negative values of $\jszz$ will 
be approximately a reflection of the presented phase boundaries with respect to the $\jszz=0$ line.
Again, the phases outside of the boundaries of the $\Gamma_5$ region are: a splayed ferromagnet (SF)
 canted from the $[100]$ cubic direction \cite{PhysRevX.1.021002}
 and the so-called $\psi_4$ \cite{Poole:2007ys} or Palmer-Chalker (PC) state.~\cite{PC1}
}
\label{fig:cMFT}
\end{figure}

The c-MFT $\psi_2/\psi_3$ phase boundary obtained for various $\jszz$ values is presented in Fig \ref{fig:cMFT}. The previous symmetry analysis in Section \ref{GLanalysis} of the $\Gamma_5$ manifold showed that a nonzero $m_z$ component is only compatible with the $\psi_2$ state. 
%As a corollary and confirmation of this expectation, we observe 
%that as  $\jszz$ is varied, the $\psi_2/\psi_3$ phase boundary position is modified. 
%Strictly in terms of the $m_z$ component of the spins on a single tetrahedron, 
%a positive $\jszz$ favors a "2-in/2-out" configuration while a 
%negative $\jszz$ favors a "4-in/4-out" configuration. 
%However, since in the $\Gamma_5$ manifold the total moment per tetrahedron is zero
%\cite{Poole:2007ys} and in particular $\sum_{i=1}^4m_i^z=0$, the "2-in/2-out" configurations are not allowed. 
%Thus, for a \emph{positive} $\jszz$, it is energetically favorable to have $m_z=0$, implying
%that the $\psi_3$ state is selected over a larger portion of the phase diagram. 
%The converse arguments follows readily for $\jszz<0$, 
%meaning that a larger $\psi_2$ region is favored for $\jszz<0$.
As a corollary and confirmation of this expectation, we observe that as $j_{zz}$ is varied, the $\psi_2/\psi_3$ phase boundary position shifts. This can be understood physically in the following way. A negative $j_{zz}$ favors a ``4-in/4-out"  
hence non-zero $m_z$ and thus the $\psi_2$ state. 
Conversely, a positive $j_{zz}$ on a single tetrahedron alone would favor a ``2-in/2-out"
spin configuration and therefore disfavor a ``4-in/4-out" configuration. However, an ordered state built with such 
``2-in/2-out" spin configuration would have a nonzero net magnetization on each tetrahedron, as well as for the whole system since we consider a ${\bm k}=0$ propagation vector, which is forbidden for the $\Gamma_5$ representation under discussion.
Therefore, the best that a positive $j_{zz}$ can do is to favor the $\psi_3$ state 
($m_z=0$) against $\psi_2$ ($m_z\neq 0$) and the boundary shifts accordingly, as shown in Fig.~\ref{fig:cMFT}. 
%Therefore, a positive $j_{zz}$ disfavors a ``4-in/4-out" $\psi_2$ state 
%and the best that the system with positive $j_{zz}$ can do 
%is to have $m_z=0$, which is a $\psi_3$ state. As a result, we expect on physical grounds, 
%and as found in the c-MFT calculations, 
%that negative $j_{zz}$ favors $\psi_2$ to the detriment of $\psi_3$ while positive $j_{zz}$ disfavors $\psi_2$ as shown in %Fig.~\ref{fig:cMFT}. 

A comment is warranted regarding the flat 
($J_{z\pm}$--independent) $\psi_2$/$\psi_3$ phase boundary 
predicted by c-MFT for $J_{zz}=0$ in Fig. \ref{fig:cMFT}.
This is to be contrasted with the E-TAP results of Fig. \ref{fig:Gamm5} 
and the $T=0$ quantum fluctuation calculations of Ref. [\onlinecite{Wong}] (see Fig. 3 therein), 
which both show a $J_{z\pm}$-dependent $\psi_2$-$\psi_3$ phase boundary for $J_{zz}=0$.
This $J_{z\pm}$ independent c-MFT result can be rationalized in the following way with 
the argument proceeding through a sequence of steps.
In the c-MFT approximation that considers a one-tetrahedron cluster, 
the $m_z$ dependence of the free-energy comes only from the $J_{zz}$ term 
because of the special role the $C_3$ symmetry plays when considering such a cluster.
The reason is that every spin on a one-tetrahedron 
cluster is coupled to three idential ``exterior''
mean fields ${\bm m}_{j,I}$ because of the ${\bm k}=0$ 
$\psi_2$ or $\psi_3$ long-range ordered
states considered.
As a result, terms such as $(\sum_i S_i^\pm)(\sum_j \zeta_{ij} m_{j,I}^z)$ vanish 
due to the $C_3$ symmetry 
that imposes the necessary form for the $\zeta_{ij}$
 bond factors.~\cite{Savary:2012uq,PhysRevLett.108.037202,PhysRevX.1.021002}
The same argument of course applies for terms of the form
$(\sum_i m_{i,I}^\pm)(\sum_j \zeta_{ij} S_j^z)$.
Consequently, the free-energy  depends on $m^z$ explicitly 
only via the $J_{zz}$ coupling since the combined dependence on $J_{z\pm}$  and $m^z$ is eliminated via the above
argument.  
When $J_{zz}=0$, all dependence of the c-MFT free-energy on $m^z$ disappears,
 and thus no-dependence on $J_{z\pm}$ can remain, as found in Fig.~\ref{fig:cMFT}.
On the other hand, when $J_{zz} \ne 0$, the c-MFT free-energy depends on $m^z$. In that case, $J_{z\pm}$ coupling 
the $S_i^z$ and $S_i^\pm$ components will, through the {\it intra-tetrahedra} 
fluctuations that are incorporated into the c-MFT calculation, renormalize the effective intra-tetrahedron  
$zz$ component sublattice susceptibility and thus the net effect of
{\it inter-tetrahedra} mean-field $(\sum_j J_{zz} m_{j,J}^z)$. 
As a result, for $J_{zz}\ne 0$, a $J_{z\pm}$ 
dependence of the $\psi_2/\psi_3$ phase boundary is observed for a single-tetrahedron c-MFT calculation (see Fig.~\ref{fig:cMFT}).

While the above argument for the $J_{z\pm}$ independence of the boundary for the $J_{zz}=0$ case applies for
a single tetrahedron cluster c-MFT, we expect the boundary for larger clusters to depend 
on $J_{z\pm}$, even for $J_{zz}=0$. In that case,
$J_{z\pm}$ may induce a $m^z$ dependence that would result in a behavior similar to 
E-TAP results for $J_{zz}=0$.
The reason is that, for larger clusters, 
the sites on the perimeter of the cluster may be coupled to 
the mean-field parameters ${\bm m}_{j,I}$ coming from less than three sites related by $C_3$ symmetry, 
and which caused the inter-tetrahedron mean-field to vanish for a 4-site tetrahedron cluster as discussed above.
Notwithstanding this caveat associated with the special symmetry of the 4-site tetrahedron cluster,
we nevertheless believe that the evolution of the boundary upon varying $J_{zz}$ 
shown in Fig. \ref{fig:cMFT} to be roughly 
qualitatively correct for $J_{zz} \ne 0$.

%======================================================================================================
%======================================================================================================

\section{Multiple transitions at $T<T_c$} 
\label{multiple_trans}

\begin{figure}[h!]
\centering
\includegraphics[scale=0.45]{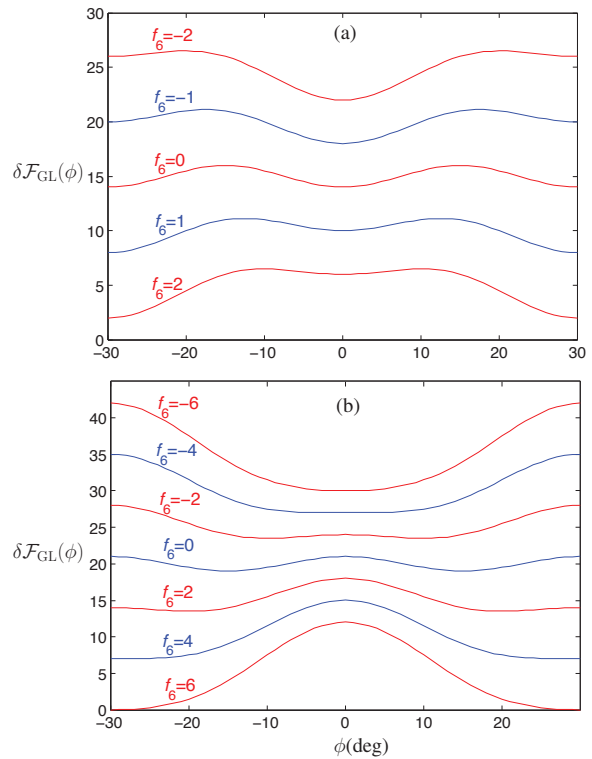}
\caption{
(Color Online) 
(a) Sequence of $\delta {\cal F}_{\mathrm {GL}}(\phi)$ with $f_{12}=-1$ and
$f_6$= -2, -1, 0, 1 and 2 (from top to bottom). There is a first-order transition at $f_6=0$.
(b) Sequence of  $\delta {\cal F}_{\mathrm {GL}}(\phi)$ with $f_{12}=1$ and $f_6$=-6, -4, -2, 0, 2 4, and 6 (from top to bottom). 
In mean-field theory, there are Ising-type second-order transitions at $f_6 = \pm 4|f_{12}|$.
}
\label{Free612}
\end{figure}

The results above, along with those of Refs.~[\onlinecite{Ludo_edge,Wong}], suggest that in some circumstances
 the state selected at $T_c$ may differ from the low-temperature phase selected by 
harmonic (classical or quantum) spin waves. 
 For example,   Ref.~[\onlinecite{Chern:2010vn}] found that, in  
a classical pyrochlore Heisenberg antiferromagnet with additional 
indirect Dzyaloshinskii-Moriya interaction, the state selected at $T\lesssim T_c$ is $\psi_2$ 
while the one selected at $T=0^+$ is $\psi_3$.
One might then ask what are the generic possibilities that may occur in $xy$ pyrochlore magnets for which the anisotropic exchange terms position them in the degenerate $\Gamma_5$ manifold.
Considering the lowest order term in $m_{xy}$, one has, as discussed above in Section \ref{GLanalysis} 
\begin{equation}
\delta {\cal F}_{\mathrm {GL}}(\phi) = f_6(T) \cos(6\phi) .
\end{equation}
 $f_6(T) \equiv 2\bar \eta_6 |m_{xy}|^6$ 
is a function of temperature, $T$, for fixed anisotropic exchange terms $J_\pm$, $J_{zz}$, $J_{z\pm}$ and $J_{\pm \pm}$, with
the sign of $f_6(T)$ dictating which state is selected at a given temperature. 
If $f_6(T=0^+)$ stays of the same sign for all $0<T<T_c$, only a single phase is realized below $T_c$.
With solely this lowest order anisotropic term in 
$\delta {\cal F}_{\mathrm {GL}}$,  the only other possibility is a first order transition between $\psi_2$ and $\psi_3$ at some temperature $T^*<T_c$ 
when  $f_6(T)$ changes sign at $T=T^*$.
However, as one gets far below $T_c$, higher order terms in $|m_{xy}|$ can become of significant magnitude in ${\cal F}_{\mathrm{GL}}$.
 One may then extend the Ginzburg-Landau free-energy by incorporating higher-order
harmonics in the anisotropy potential, as 
\begin{equation}
\delta {\cal F}_{\mathrm {GL}}(\phi) = \sum_{n=1} 2\bar \eta_{6n}(T) |m_{xy}|^{6n}\cos(6n\phi).
\end{equation}
Keeping only the two lowest order terms for illustration purposes, we have 
\begin{equation}
\delta {\cal F}_{\mathrm {GL}}(\phi) =  f_{6}(T) \cos(6\phi) +  f_{12}(T) \cos(12\phi) , 
\end{equation}
where $f_{6n}\equiv 2\bar \eta_{6n} |m_{xy}|^{6n}$.
For simplicity,  consider first $f_{12}=-1$
%, taking $|m_{xy}|=1$.
%Thermal fluctuations decreasing $|m_{xy}|$ below $|m_{xy}|=1$ 
%would merely amount to renormalize the $f_{6n}$ and shift phase transition boundaries.
As $f_6(T)$ varies from $f_6(T)<0$ to $f_6(T)>0$, a first order transition occurs when the minimum
of $\delta {\cal F}_{\mathrm {GL}}$ shifts discontinuously,
 in a first order transition, from  $\phi=0$ ($\psi_2$ state) to $\phi=\pi/6$ ($\psi_3$ state)
(see Fig. \ref{Free612}a). 
A more interesting behavior is in principle possible. 
Consider now $f_{12}=+1$.
In that case, second order (Ising-type) transitions occur at $f_6=\pm 4 |f_{12}|$.
 For example, for $f_6 < -4$, the minimum is at $\phi=0$ (a $\psi_2$ state), 
and a second order transition to a $f_6$-dependent angle $\phi$ occurs at $f_6=-4$. 
As $f_6$ continues growing and become less negative (at fixed $f_{12}=+1$), 
the magnetic moment orientation $\phi$ continues increasing
untill another second order transition to the $\psi_3$ state occurs at $f_6 = +4$  
(see Fig. \ref{Free612}b).
While the direct $\psi_2$ to $\psi_3$ transition observed in the pyrochlore Heisenberg 
antiferromagnet with indirect Dzyaloshinskii-Moriya interaction \cite{Chern:2010vn} suggests that this system belongs to the first case, the second
scenario is perhaps not excluded in more complex models in which both temperature 
and applied magnetic field are varied simultaneously, 
or when the magnetic Er$^{3+}$ sites are diluted by non-magnetic ions.~\cite{Zhito_dil,McClarty_dil}

In the above discussion, we have implicitly assumed that the anisotropic terms  
$[m_z(m_{xy})^3 + {\rm c.c.}]^k$ and $[(m_{xy})^6 + {\rm c.c.}]^l$, which give terms of the form
$\vert m_{xy}\vert^{6n}\cos(6n\phi)$ in the Ginzburg-Landau theory once $m_z$ has been eliminated (see discussion in Section \ref{GLanalysis}), 
are generated by thermal fluctuations beyond the 
Ginzburg-Landau free-energy derived by a s-MFT treatment of the microscopic model
${\cal H}$ in Eq.~(\ref{eq:hami}).
However, it is in principle possible that virtual crystal field excitations (VCFE)  
\cite{Petit_ObD,McClarty_Er2Ti2O7,Hamid_TTO,Hamid_Paul_Mcihel} mediated by the bare multipolar
 interactions between the rare-earth ions would generate \cite{Petit_ObD,Curnoe2008} an effective pseudospin-$1/2$ Hamiltonian more complex than the one
given by Eqs.~(\ref{eq:hami}a), (\ref{eq:hami}b) and (\ref{eq:hami}c) and 
involve multispin (e.g. ring-exchange like) interactions capable of lifting 
degeneracy at the classical level without invoking order-by-disorder.~\cite{Zhitomirsky:2012fk,Savary:2012uq}
 However, one naively expects those interactions generated by VCFE to be signigicantly smaller compared with
the $J_{\pm}$, $J_{\pm\pm}$, $J_{z\pm}$ and $J_{zz}$ of Eq.~(\ref{eq:hami}a),
as stated in Ref.~\onlinecite{Savary:2012uq}.
Whether those terms are truly inefficient in competing with order-by-disorder, once the
commonly large prefactors of combinatoric origin arising in high-order perturbation theory
have been taken into account, must await a detailed calculation.

\section{Discussion}\label{sec:disc}

In this work, we first used an extended TAP (E-TAP) method to analytically study the problem of 
order-by-disorder (ObD) near the critical temperature
of a general three-dimensional $xy$ antiferromagnetic model on the pyrochlore lattice
% in a system of 3D $xy$ spins on the pyrochlore lattice
with the Hamiltonian of Eq. (\ref{eq:hami}) with $J_{zz}=0$. 
We focused on the $\Gamma_5$ manifold which is $U(1)$ degenerate at the standard mean-field theory (s-MFT) level.
The fluctuations corrections to the free-energy beyond s-MFT were organized as an expansion in powers of 
the inverse temperature, $\beta$,  and to  lowest orders in the coupling constants 
$j_{\pm\pm} = J_{\pm\pm}/J_\pm$, $j_{z\pm} = J_{z\pm}/J_\pm$ and the on-site magnetization $\bm m$.
 We established that in different parts of the $\Gamma_5$ region, the $\psi_2$ and $\psi_3$ states 
are the only minima of the free-energy selected by fluctuations up to the lowest order $U(1)$ symmetry-breaking terms considered.
The phase boundary between $\psi_2$ and $\psi_3$ 
can then be obtained by finding the real roots of  Eq. (\ref{eq:mainC}) in terms of $\jpp/\jp$ and $\jzp/\jp$. 
We also numerically studied the ObD mechanism in the 3D-$xy$ pyrochlore system a using cluster mean-field theory (c-MFT). 
Using this method, we obtained a phase boundary between $\psi_2$ and $\psi_3$ states for a variety of $\jzz$ values shown in Fig. \ref{fig:cMFT}.

Using a Ginzburg-Landau theory,  along with E-TAP and c-MFT calculations, we predict that for a state ordered in the local 
$xy$ plane  at the s-MFT level, 
fluctuations can induce a small out-of-$xy$-plane $m_z$ component of the on-site magnetization.
 This fluctuation-induced local $z$ component of the magnetization is only compatible with the $\psi_2$ state. 
We expect the size of this moment to be small, since it is proportional to $|m_{xy}|^3 \ll 1$ for $T \lesssim T_c^{\mathrm{MF}}$ (see Eq. (\ref{mz-GL})). 
Yet,  on the basis of the c-MFT results, we might  anticipate that 
nonzero $m_z$ has an important
 effect on the phase boundary between the $\psi_2$ and the 
$\psi_3$ states for $J_{zz} \neq 0$, as illustrated in Fig. \ref{fig:cMFT} (see also the caption of the figure).

By considering the single phase boundary 
between the $\psi_2$ and $\psi_3$ states for $T\lesssim T_c$ 
 along with the multiple $\psi_2$-$\psi_3$ boundaries found at $T=0^+$ in Ref. [\onlinecite{Wong}], 
for $\jspp \ll 1 $ and $\jszp \ll 1$, the regime for which the perturbative E-TAP solutions above apply, 
one may expect to find multiple phase transitions between $\psi_2$ and $\psi_3$ states upon decreasing temperature from $T\lesssim T_c$ to $T\sim 0^+$.
While possible in principle and found in a numerical study,~\cite{Chern_DM}
such multiple transitions in real $xy$ pyrochlore magnetic materials have not yet been reported.
Since in  the E-TAP calculations the $\jspp$ and $\jszp$ couplings were treated perturbatively,  
the E-TAP phase boundary in Eq. (\ref{root}) is valid for small $\jspp$ and $\jszp$ and would thus need to be  modified for non-perturbative 
$\jspp$ and $\jszp$, that is, closer to the boundaries of the $\Gamma_5$ region with  the splayed-ferromagnet (SF) phase
 (see Fig.~(\ref{fig:Gamm5})).
A case in point is the Heisenberg pyrochlore antiferromagnet with indirect Dzyaloshinskii-Moriya interaction.~\cite{Chern_DM}
This model displays a transition at $T_{c}^{+}$ from a paramagnetic state to $\psi_2$, followed at $T_{c}^{-}<T_{c}^{+}$ by a transition from
$\psi_2$ to $\psi_3$,\cite{Chern_DM}   and 
for which the corresponding anisotropic exchange $J_\pm$, $J_{\pm\pm}$, $J_{z\pm}$ and $J_{zz}$ are such 
that this model lives \emph{right on} the $\Gamma_5$ to SF classical phase boundary.\cite{Wong}

The difference between the phase boundaries obtained by E-TAP and c-MFT for $\jzz=0$ (see Figs.~\ref{fig:Gamm5} and \ref{fig:cMFT})
arises from the different range of fluctuations considered in these two methods. 
In c-MFT, because of the exact diagonalization of the Hamiltonian on a single 
tetrahedron and the s-MFT treatment of the inter-tetrahedra couplings,
the fluctuation corrections considered are short-ranged (limited to the nearest-neighbors). 
However in the E-TAP method, by considering the higher order terms in the $\beta$ expansion,
specifically $\beta^4$ and beyond where
 larger (more extended) diagrams appear in the calculation (see Appendix \ref{apA}, Section \ref{sec:diagrams}), 
fluctuations beyond nearest-neighbors are included. 
It would be of  interest to benchmark these arguments by performing c-MFT calculations for larger clusters.

The E-TAP corrections to the s-MFT free-energy could also be computed for the $\jzz\neq 0$ case.  
The procedure for considering $J_{zz}\neq 0$  is conceptually the same as for the $J_{zz}=0$ case for which, 
 using the E-TAP method, we calculated the terms in the high temperature expansion of Eq. (\ref{Gibbs1}) 
that contribute to the degeneracy-lifting $\cos(3\phi)$ and $\cos(6\phi)$ terms in the free-energy.  
To obtain these terms to the lowest order in the coupling constants $\jpp$, $\jzp$ and $\jzz$,
 one would need to consider the high temperature expansion of $G$ up to the seventh order in $\beta$ and
calculate the degeneracy lifting terms following the power-counting prescription explained above Eq. (\ref{eq:g0}). 
However, the number of terms of the form  $\jszz^l\jspp^m\jszp^n$,
with fixed $l+m+n$ value $l$, $m$ and $n$ positive integers, 
and thus the number of diagrams to be computed,
proliferate significantly as to make this calculation a significant undertaking. 
The $\jzz\neq 0$ case thus stands on its own as an independent future study.

% to explore the effect of fluctuations with length scale beyond a single tetrahedron.

Finally, we remark that variants of the E-TAP method presented in this work
could be applied to models other than the pyrochlore structure to analytically investigate the role of fluctuations
 beyond s-MFT in selecting a specific long-range ordered state close to the critical temperature. One example is the case of Heisenberg spins on face centered cubic (FCC) 
lattice interacting via magnetostatic dipole-dipole interaction. \cite{Roser_Corruccini,Bouchaud}
 Generally speaking, one might expect that a consideration of an E-TAP analytical description 
of fluctuations at the microscopic level may help shed light on the role of individual 
symmetry-allowed interactions in the ordered state selection
near the critical temperature over a broad range of  frustrated spins models
for which an accidental degeneracy exists in a standard mean-field theory description.

%=========================================================================================
%=========================================================================================

\begin{acknowledgements}

 We acknowledge Paul McClarty, Ludovic Jaubert, Jaan Oitmaa, Rajiv Singh and 
Anson Wong for useful discussions and collaborations related to this work. 
We also thank  Antoine Georges for a useful discussion.
We are grateful to
Peter Holdsworth and Natalia Perkins for constructive comments and suggestions on 
an earlier version of the manuscript. 
This work was supported by the NSERC of Canada, the Canada Research Chair program (M.G., Tier 1) 
and by the Perimeter Institute (PI) for Theoretical Physics. 
Research at PI is supported by the Government of Canada through Industry Canada and by the
Province of Ontario through the Ministry of Economic Development \& Innovation.

\end{acknowledgements}

\appendix
%\onecolumngrid
%=========================================================================================
%========================================================================================
\section{$\psi_2$ and $\psi_3$ states
 \label{apPsi23}}

The $\psi_2$ states have spin configurations with a ${\bm k}=0$ ordering wave vector and with the following spin orientations on a tetrahedron expressed in the global (Cartesian) reference frame
 \begin{subequations}\label{psi2}
\begin{eqnarray}
\hat{e}_0^1&=&\frac{1}{\sqrt{6}}( -1, -1, 2)\\
\hat{e}_1^1&=&\frac{1}{\sqrt{6}}(1, 1, 2)\\
\hat{e}_2^1&=&\frac{1}{\sqrt{6}}(1, -1, -2 )\\
\hat{e}_3^1&=&\frac{1}{\sqrt{6}}( -1, 1, -2)
\end{eqnarray}
\end{subequations}
Here the subscripts correspond to the four sublattice labels in the pyrochlore lattice and the superscripts refer to different symmetry-related $\psi_2$ states. 
$\{\hat{e}_i^2\}$ and $\{\hat{e}_i^3\}$, $i=0,\cdots, 3$, can be obtained from Eq. (\ref{psi2}) 
by $C_3$ ($\pi/3$) rotations with respect to the $\langle 111 \rangle$ directions. 
$\psi_3$ states can be obtained from the $\psi_2$ ones, by a $\pi/6$ rotation of each spin about its local $[111]$ axis.
%=========================================================================================
%========================================================================================

\section{Symmetry groups of $\psi_2$ and $\psi_3$ states}
\label{ap:symm}

The symmetries of either $\psi_i$ state ($i=2,3$) form a group known as the little group of the corresponding state. The generators of the $\psi_3$ little group include the $C_2$ rotation by $\pi$ about one of the cubic $x$, $y$ or $z$ axes (depending on the particular $\psi_3$ state), the improper rotation by $\pi/2$ ($S_4$) about the same cubic axis,  and two plane reflections $\sigma_1$ and $\sigma_2$ with respect to planes spanned by the cubic axis and one of the two tetrahedron bonds perpendicular to the cubic axis.  The generators of the little group of $\psi_2$ states include $\tau S_4$, $\tau\sigma_1$ and $\tau\sigma_2$ where $\tau$ is the time-reversal symmetry operation. 

Considering the action of the symmetry operations in the $\psi_2$'s and $\psi_3$'s little groups on the possible configurations with 
a finite onsite $m_z$ moments on a single tetrahedron 
(i.e. all-in/all-out, 2-in/2-out, 3-in/1-out and 1-in/3-out where ``in'' and ``out'' indicates 
whether $m_{i,z}$ is positive or negative on site $i$), 
only all-in-all-out configuration is invariant under the symmetry operations of the $\psi_2$'s little group while none of the above configurations are invariant under the $\psi_3$'s little group symmetry operations.  As a result,  only the $\psi_2$ state can possess a finite onsite $m_z$ moment which 
displays an all-in-all-out configuration on a single tetrahedron and therefore does not produce a net magnetic moment on a tetrahedron.
%=========================================================================================
%========================================================================================

\section{Extended TAP Method (E-TAP) \label{apA}}

In this Appendix, we first derive the general form of the E-TAP corrections up to sixth order in $\beta$. Next, we illustrate our method for calculating the terms in the $\beta$ expansion of the Gibbs free-energy, $G$, of Eq.~(\ref{Gibbs1})  by focusing on two calculation examples. 
Finally, we discuss the general properties of the diagrammatic approach employed to carry out the computation of the various terms entering the E-TAP calculation.

\subsection{Derivation}
 \label{deriv}

The E-TAP method is based on a perturbative expansion of the Gibbs free-energy as a function of inverse temperature, $\beta$, beyond that given by s-MFT. 
There are several ways of deriving these corrections. \cite{naiveTAP} 
Here, we employ and extend the method presented by Georges and Yedidia \cite{Yadidah} for Ising spin glass which provides the corrections due to fluctuations about the s-MFT solution, order by order in  $\beta$. 
 In this part of the Appendix, we focus on a temperature regime close to the mean-field transition temperature, 
$T \lesssim T_c^{\mathrm{MF}}$, and derive the E-TAP corrections for classical Heisenberg spins of fixed length $|\bm S|=1/2$.

%We consider the Gibbs free energy in Eq.(\ref{Gibbs1}). 
The Hamiltonian in Eq. (\ref{eq:hami}), $\mathcal{H}$, can be written as:
\begin{equation}
\label{Ham1}
\mathcal{H}=\sum_{i<j}S_i^{\mu}J_{ij}^{\mu\nu} S_j^{\nu},
\end{equation}
 Greek labels represent Cartesian coordinates and implicit summation over repeated superscripts is used.
A Taylor series expansion of Eq. (\ref{Gibbs1}) in powers of $\beta$ reads:
\begin{eqnarray}
\label{expansion1-A}
 G(\beta)&=& \frac{1}{\beta}\Big( \tilde{G}(\beta)\big|_{\beta=0} +\frac{\partial  \tilde{G}(\beta)}{\partial\beta}\Big|_{\beta=0}\beta \\ \nonumber
 &&+\frac{1}{2!}\frac{\partial^2 \tilde{G}(\beta)}{\partial\beta^2}\Big|_{\beta=0}\beta^2\cdots\Big),
 \end{eqnarray}
where  $\tilde{G}(\beta)=\beta G(\beta)$. We define,
\begin{eqnarray}
\label{udef}
U&\equiv&\frac{1}{2}\sum_{ij}\delta S_i^{\mu}J_{ij}^{\mu\nu}\delta S_j^{\nu} 
\end{eqnarray}
where
\begin{eqnarray}
\label{deltadef}
\delta S_j^{\nu}&\equiv&S_j^{\nu}- m_j^{\nu} 
\end{eqnarray}
are spin fluctuations about the s-MFT solution.
As shown in Ref. [\onlinecite{Yadidah}], the derivatives of $ \tilde{G}(\beta)$ with respect to $\beta$ can be evaluated in
 terms of expectation values of powers of $U$ and $T_n$, that read:
\begin{widetext}
\begin{subequations}\label{TAPP}
\begin{eqnarray}
\frac{\partial(\beta G(\beta))}{\partial\beta}&=&\langle \mathcal{H}\rangle, \label{TAPS1} \\
\frac{\partial^2 (\beta G(\beta))}{\partial\beta^2}&=& -\langle U^2\rangle, \label{TAPS2} \\
\frac{\partial^3 (\beta G(\beta))}{\partial\beta^3}&=& \langle U^3 \rangle, \label{TAPS3}\\
\frac{\partial^4 (\beta G(\beta))}{\partial\beta^4}&=&- \langle U^4\rangle + 3\langle U^2\rangle^2-3\langle U^2T_2\rangle, \label{TAPS4} \\
\frac{\partial^5 (\beta G(\beta))}{\partial\beta^5}&=& \langle U^5\rangle-10\langle U^2\rangle\langle U^3\rangle-3\langle U^2T_3\rangle+7\langle U^3T_2\rangle + 6\langle UT_2^2\rangle, \label{TAPS5}\\
\frac{\partial^6 (\beta G(\beta))}{\partial\beta^6}&=& -\langle U^6\rangle+15\langle U^4 \rangle\langle U^2\rangle + 10\langle U^3\rangle^2 - 30 \langle U^2\rangle^3 -12 \langle U^4T_2\rangle\\ \nonumber 
&&+10\langle U^3T_3\rangle -3\langle U^2T_4\rangle-27\langle U^2T_2^2\rangle+ 18\langle UT_2T_3\rangle \\ \nonumber
&&-6\langle UT_3\rangle\langle U^2\rangle + 51\langle U^2\rangle\langle U^2T_2\rangle + 6\langle U^2\rangle\langle T_2^2\rangle-6\langle T_2^3\rangle, \label{TAPS6}
\end{eqnarray}
\end{subequations}
\end{widetext}
and where $T_n$ is defined as 
\begin{equation}
\label{Tdef}
T_n \equiv \sum_i \frac{\partial^n\bm \lambda_i}{\partial\beta^n}\cdot\delta\mathbfit S_i.
\end{equation} 
%Recalling the relation between $\bm \lambda_i$ and the local mean-field $\bm h_i$, 
Since $\partial {\bm \lambda}/\partial \beta |_{\beta=0} = \bm h_i$\cite{Yadidah} where $h_i^{\mu}=\sum_{j,\nu}J_{ij}^{\mu\nu}m_j^{\nu}$,  the terms involving
involving $T_n$ come from considering fluctuations of the local mean-field.
 On the other hand, the $U^n$ terms take into accont
the fluctuations of the on-site magnetization  
(see Eqs. (\ref{udef}, \ref{deltadef}))
within the ensemble set by the s-MFT solution.
%while $T_n$ incorporates the fluctuations in the mean field.
The $\langle \cdots \rangle$ above denotes a thermal average. For a general observable $O$, $\langle O\rangle$ is given by:
\begin{equation}
\langle O\rangle=\frac{Tr[O \exp\big(-\beta H+ \sum_i \bm \lambda_i \cdot (\mathbfit S_i-\mathbfit m_i)\big) ]}{Tr[\exp\big(-\beta H+ \sum_i \bm \lambda_i \cdot (\mathbfit S_i-\mathbfit m_i)\big)]}.
\label{DefO}
\end{equation}
The first two terms in Eq. (\ref{expansion1-A}) correspond to the s-MFT free-energy while the higher order terms  in $\beta$ provide corrections beyond s-MFT.
 Calculating the expectation value of powers of $U$ at $\beta=0$ reduces to the evaluation of mean-field averages of the  form:
\begin{equation}
\label{nnzero}
 \langle \delta S^{\alpha_1}_{i_1}\delta S^{\alpha_2}_{i_2}\cdots\delta S^{\alpha_n}_{i_n}\rangle_{\mathrm {MF}},
 \end{equation}
 where $i_n$ represents the site label and $\alpha_n$ represents a Cartesian coordinate. $n$ is the number of $\delta S$ factors in Eq. (\ref{nnzero}).
From now on, we drop the $\mathrm {MF}$ subscript for compactness of notation.
% in Eqs.~(\ref{TAPS1}-\ref{DefO}), we dropped the $\mathrm {MF}$ subscript for %compactness of notation.
For $n=1$, the expectation value in Eq. (\ref{nnzero}) is zero due to the relation  $\mathbfit m_i=\langle \mathbfit S_i \rangle$. For $ n \geq 2 $, however, Eq. (\ref{nnzero}) is nonzero only if there is no site label that appears only once. For example, averages of the following form have a nonzero contribution:
 \begin{equation}
 \label{nnzero-2}
 \langle \delta S^{\alpha_1}_{i}\delta S^{\alpha_2}_{i}\delta S^{\alpha_3}_{j}\delta S^{\alpha_4}_{j}\delta S^{\alpha_5}_{j}\rangle= \langle \delta S^{\alpha_1}_{i}\delta S^{\alpha_2}_{i}\rangle\langle\delta S^{\alpha_3}_{j}\delta S^{\alpha_4}_{j}\delta S^{\alpha_5}_{j}\rangle
 \end{equation}
 
  The expectation values above can be calculated using the self-consistent s-MFT equations for 3-component classical spins which are given by the Langevin function
  \begin{equation}
 \label{slfMF1}
\mathbfit m_i = \frac{\bm \lambda_i}{|\bm \lambda_i|}\Big[\coth(|\bm \lambda_i|)-\frac{1}{|\bm \lambda_i|}\Big].
\end{equation}
%and for the case of  spin-$1/2$ reads
%\begin{equation}
% \label{slfMF1}
%\mathbfit m_i = -\frac{\mathbfit h_i}{2|\mathbfit h_i|}\tanh(\frac{|\mathbfit h_i|}{2}).
%\end{equation} 
Consequently 
 \begin{equation}
 \label{chi1}
 \langle\delta S^{\alpha}_i\delta S^{\beta}_i\rangle = \frac{\partial m^{\alpha}_i}{\partial \lambda^{\beta}_i} \equiv \chi_i^{\alpha\beta} ,
 \end{equation}
 \begin{equation}
 \label{chi2}
 \langle\delta S^{\alpha}_i\delta S^{\beta}_i\delta S^{\gamma}_i\rangle = \frac{\partial\chi_i^{\alpha\beta}}{\partial \lambda^{\gamma}_i},
 \end{equation}
and in general
\begin{equation}
\label{chin}
 \langle \delta S^{\alpha_1}_{i}\delta S^{\alpha_2}_{i}\cdots\delta S^{\alpha_n}_{i}\rangle=\frac{\partial \langle \delta S^{\alpha_1}_{i}\delta S^{\alpha_2}_{i}\cdots\delta S^{\alpha_{n-1}}_{i}\rangle}{\partial \lambda_i^{\alpha_n}}.
 \end{equation} 

Since we are interested in a temperature range close to $T_c$,  Eq. (\ref{slfMF1}) can be expanded for small $|\bm \lambda_i|$:
\begin{equation}
\label{exp-selfi}
m_i^{\alpha}=\frac{\lambda_i^{\alpha}}{3}[1- \frac{(|\bm \lambda_i|)^2}{15} + \frac{2(|\bm \lambda_i|)^4}{315} + \cdots ],
\end{equation}
 The expansions of Eqs. (\ref{chi1},\ref{chi2}, \ref{chin}) at $T\lesssim T_c$ 
can be computed by differentiating Eq. (\ref{exp-selfi}) with respect to different components of the vector $\bm \lambda_i$ Lagrange multipler.

In the rest of this section, we focus on the specific
 problem of the s-MFT $U(1)$ degeneracy of the $\Gamma_5$ manifold
displayed by the Hamiltonian of Eq. (\ref{eq:hami}) in a wide range of the 
$\{J_\pm, J_{\pm\pm}, J_{z\pm},J_{zz}\}$ anisotropic exchange parameters.\cite{Wong} 
Below, we calculate the degeneracy-lifting contributions from $\langle U^2 \rangle$ and $\langle U^2T_2\rangle$ in Eqs. (\ref{TAPS2}, \ref{TAPS4}) to demonstrate the general idea of the method. Higher order terms can be calculated using a similar procedure.
%Below we demonstrate our method by calculating  terms in Eq.(\ref{TAPS4}) as two examples.  

\subsection{$\langle U^2 \rangle$ Calculation} 
 Using Eq. (\ref{udef}), we have
\begin{equation}
\label{U2}
\langle U^2 \rangle= \frac{1}{2^2}\sum_{\langle i_1, j_1\rangle}\sum_{\langle i_2,j_2\rangle}J^{\alpha_1\beta_1}_{i_1j_1}J^{\alpha_2\beta_2}_{i_2j_2}\langle\delta S_{i_1}^{\alpha_1}\delta S_{j_1}^{\beta_1}\delta S_{i_2}^{\alpha_2}\delta S_{j_2}^{\beta_2}\rangle,
\end{equation}
where the summations are performed over lattice sites and summation is implied for Greek superscripts. 
To proceed, we need to specify all the possible pairings of the factors $\delta S_{i_t}$, $\delta S_{j_t}$ ($t=1,2$) in the $\langle\delta S_{i_1}^{\alpha_1}\delta S_{j_1}^{\beta_1}\delta S_{i_2}^{\alpha_2}\delta S_{j_2}^{\beta_2}\rangle$.

Considering the description provided above Eq. (\ref{nnzero-2}), 
the only nonzero pairings of site indices in Eq. (\ref{U2}) 
are $i_1=i_2$, $j_1=j_2$ and $i_1=j_2$, $j_1=i_2$, 
with the constraint $i_1\neq j_1$ and $i_2\neq j_2$ imposed by the Hamiltonian; $J_{ii}^{\alpha\beta}=0$ for all $\alpha$ and $\beta$. This leads to:
\begin{equation}
\label{U2-2}
\langle U^2 \rangle= \frac{1}{2}\sum_{\langle i,j\rangle}J^{\alpha_1\beta_1}_{ij}J^{\alpha_2\beta_2}_{ij}\langle\delta S_{i}^{\alpha_1}\delta S_{i}^{\alpha_2}\rangle\langle\delta S_{j}^{\beta_1}\delta S_{j}^{\beta_2}\rangle.
\end{equation}
 %We note that by substituting Eq. (\ref{U2-2}) in Eq. (\ref{TAPS2}) and using Eq. (\ref{chi1}), we recover  Eq. (\ref{TAP2}). 

\begin{figure}
\begin{center}
\includegraphics[width=7 cm]{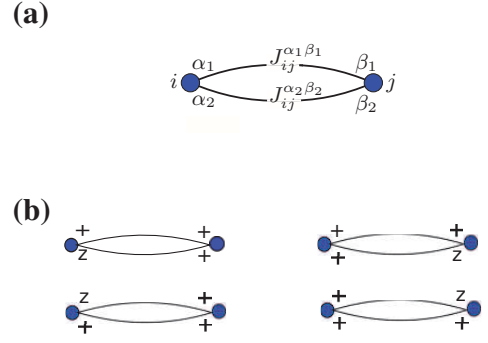}
%\put(-200,115){{\bf a)}}
%\put(-200,55){{\bf b)}}
\caption{{\bf a)} Diagrammatic representation of the term $J^{\alpha_1\beta_1}_{ij}J^{\alpha_2\beta_2}_{ij}\langle\delta S_{i}^{\alpha_1}\delta S_{i}^{\alpha_2}\rangle\langle\delta S_{j}^{\beta_1}\delta S_{j}^{\beta_2}\rangle$ in Eq. (\ref{U2-2}). {\bf b)} All diagrams corresponding to $\alpha_i, \beta_i=z,+$, $i=1,2$ with $z$ appearing only once among the Greek superscripts. The same number of diagrams are present for the case where $\alpha_i, \beta_i=z,-$. Again $z$ appears only once.\label{diagU2}}
\end{center}
\end{figure}

The computational complexity of the method requires the usage of a diagrammatic approach, which we now introduce by considering, 
for example,
the calculation of Eq. (\ref{U2-2}). In this equation, for a given $i$, $j$, $\alpha_t$ and $\beta_t$ with $t=1,2$, $J^{\alpha_1\beta_1}_{ij}J^{\alpha_2\beta_2}_{ij}\langle\delta S_{i}^{\alpha_1}\delta S_{i}^{\alpha_2}\rangle\langle\delta S_{j}^{\beta_1}\delta S_{j}^{\beta_2}\rangle$ can be represented by a diagram of the form illustrated in Fig. \ref{diagU2}a. In this figure, the vertex labels match the labels of the lattice sites that the diagrams cover. Each vertex represents an average of the form $\langle \delta S_i^{\alpha_1}\cdots\delta S_i^{\alpha_t}\rangle$ where $t$ is the number of bonds (solid lines) connected to that vertex. In the case of Eq. (\ref{U2-2}) as indicated in Fig. \ref{diagU2}a, the averages are $\langle\delta S_{i}^{\alpha_1}\delta S_{i}^{\alpha_2}\rangle$ and $\langle\delta S_{j}^{\beta_1}\delta S_{j}^{\beta_2}\rangle$. The bonds represent the elements of the coupling matrix, $J_{ij}^{\alpha\beta}$.
% For a given $i$ and $j$ pair, all the terms in the implicit sum can be obtained by assigning $z$, $+$ and $-$ to the Greek superscripts. 
It is straightforward combinatorics to take into account only the terms that contribute to the degeneracy-lifting factors, $\cos(3\phi)$ or $\cos(6\phi)$. We next proceed to demonstrate this point. 

 Based on the symmetry analysis of Section \ref{GLanalysis}, 
we note that since $\langle U^2 \rangle$ involves  four powers of $\delta S_i^{\alpha_t}$.
$\langle U^2 \rangle$  can only contribute degeneracy-lifting terms of the general form  $\propto m_z|m_{xy}|^3\cos(3\phi)$ and
 where the proportionality factor is generated by the anisotropic couplings $\jspp$ and $\jszp$ in Eq. (\ref{h1}). We recall that $\jspp$ and $\jszp$ each contributes a factor $e^{\pm2i\phi}$ and $e^{\pm i\phi}$ in the power counting method. The $-i\phi$ and $-2i\phi$ corresponds to presence of  $\delta S^-$ and $(\delta S^-)^2$ in Eq. (\ref{U2-2}), respectively, which in turn, implies the presence of $J_{ij}^{--}\equiv\jspp\gamma_{ij}^{\ast}$ or $J_{ij}^{z-}\equiv\jszp\zeta_{ij}^{\ast}$ matrix elements in  $J^{\alpha_1\beta_1}_{ij}J^{\alpha_2\beta_2}_{ij}\langle\delta S_{i}^{\alpha_1}\delta S_{i}^{\alpha_2}\rangle\langle\delta S_{j}^{\beta_1}\delta S_{j}^{\beta_2}\rangle$. So for a given $i$ and $j$, only the following combination of terms can generate $\cos(3\phi)$:
 \begin{eqnarray}
   & &  J^{++}_{ij}J^{z+}_{ij}\langle\delta S_{i}^{+}\delta S_{i}^{+}\rangle\langle\delta S_{j}^{z}\delta S_{j}^{+}\rangle + \nonumber \\ 
   & & J^{--}_{ij}J^{z-}_{ij}\langle\delta S_{i}^{-}\delta S_{i}^{-}\rangle\langle\delta S_{j}^{z}\delta S_{j}^{-}\rangle,
 \end{eqnarray}
 where the first term generates the $e^{3i\phi}$ contribution to $\cos(3\phi)$ while the second term generates $e^{-3i\phi}$. All possible ways of generating the factor $e^{3i\phi}$ are illustrated in Fig. \ref{diagU2}b, which is the same as the number of ways of generating $e^{-3i\phi}$. Ultimately, Eq. (\ref{U2-2}) can be rewritten as:
%the degeneracy lifting term of the form $\jspp\jszpin Eq. (\ref{U2-2}) is presented in Fig. \ref{diagU2}b and the Eq. (\ref{U2-2}) will be reduced to
 \begin{equation}
 \label{U2-final1}
 \langle U^2 \rangle= 2\jspp\jszp\Big(\langle \delta S^{z}\delta S^{+}\rangle\langle\delta S^{+}\delta S^{+}\rangle\sum_{ij}\gamma_{ij}\zeta_{ij}+ \mathrm{h.c.}\Big).
 \end{equation}
 Based on Eqs. (\ref{chi1}, \ref{exp-selfi}) and recalling from Section \ref{sec:mod} that $\bm m$ and $\phi$ do not require site indices in the $\Gamma_5$ manifold, we have dropped the site labels of $\delta S$'s in Eq. (\ref{U2-2}) when writing Eq. (\ref{U2-final1}). The lattice sum, $\sum_{ij}\gamma_{ij}\zeta_{ij}$, can be carried out using a computer program for different lattice sizes with linear dimension $L$.
Up to sixth order in the $\beta$ expansion, the lattice sums \emph{per site} for different terms in Eqs. (\ref{TAPP}) are independent of $L$ for $L \geq 2$ . Accordingly, we have
\begin{equation}
\frac{1}{N}\sum_{ij}\gamma_{ij}\zeta_{ij}=-6.
\end{equation}
 
\begin{figure}
\begin{center}
\includegraphics[width=8.25 cm]{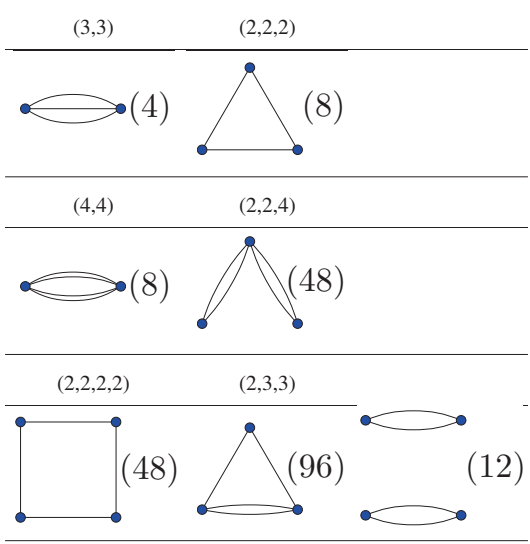}
\caption{ For details, see the text in  Appendix \ref{sec:diagrams}. We also note that the contribution of disconnected diagrams (e.g. the diagram at the bottom right corner) in Eq. (\ref{TAPP}) adds up to zero.
\label{tab:graph}}
\end{center}
\end{figure}

 Using Eqs. (\ref{chi1}, \ref{exp-selfi} ), the averages in Eq. (\ref{U2-final1}) can be written as
 \begin{eqnarray}
 \langle \delta S^{z}\delta S^{+}\rangle&=& \frac{18}{45}m_zm_++\cdots \\
 \langle \delta S^{+}\delta S^{+}\rangle&=& \frac{18}{45}m_+^2+\cdots
 \end{eqnarray}
 where to the lowest order of interest in $\bm m$ which, in this case, is the fourth order, we kept $\lambda^{\alpha} \simeq 3m^{\alpha}$ and neglect all higher order terms. Finally Eq. (\ref{U2-final1}) gives
 
 \begin{eqnarray}
\label{fin-sample}
  \langle U^2 \rangle/N & = & (-0.96m_zm_+^3+\mathrm{h.c.}) \jspp\jszp 		\nonumber\\
  				& = & -1.92 \jspp\jszp m_z|m_{xy}|^3\cos(3\phi). 
  \end{eqnarray}
 We note that Eq. (\ref{fin-sample}) contributes to  
$\omega$ in Eq. (\ref{GLfin2}) which, in turn, is necessary to evaluate  $\bar \eta_6(\jspp, \jszp)$ in Eq. (\ref{freefinal}). 
 All other terms in Eq. (\ref{TAPP}) involving solely powers of $U$ and no $T_n$ terms
can be calculated in a similar way. The number of diagrams increases as one considers higher order terms in the $\beta$ expansion of Eq. (\ref{Gibbs1}).  As a result, finding the number and type of nonzero average of the form of Eq. (\ref{nnzero}) is most easily done using a computer program. The details of this type of calculations are presented in Appendix \ref{sec:diagrams}. Some of the diagrams that appear at higher order in the $\beta$ expansion are illustrated in Fig. \ref{tab:graph}. 

\subsection{$\langle U^2T_2 \rangle$ Calculation}\label{sec:U2T2}

\begin{figure}[h]
\begin{center}
\includegraphics[width=2.25 cm]{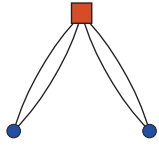}
%\put(-80,50){\large{b)}}
%\put(-170,50){\large{a)}}
\caption{ 
 ``Fused'' diagram corresponding to $\langle U^2T_2\rangle$. The vertex represented by the (red) square represents the following: $ \langle\delta S_{k}^{\alpha_3}\delta S_{k}^{\beta_1}\delta S_{k}^{\beta_2}\rangle \langle\delta S_{k}^{\alpha_3}\delta S_{k}^{\gamma_1}\delta S_{k}^{\gamma_2}\rangle$. 
 \label{graph6}}
\end{center}
\end{figure}

 Due to the presence of $T_n$ in Eq. (\ref{Tdef}), in this case $n=2$, averages that involve $T_n$
need to be carried out slightly differently in comparison with averages containing only powers of $U$. The difference comes from the presence of the factor $\partial^n \bm \lambda_i/\partial\beta^n$ in Eq. (\ref{Tdef}). Considering Eq. (\ref{TAPS1})
and the relation 
$\partial {\bm \lambda}/\partial \beta |_{\beta=0}  = \bm h_i$, \cite{Yadidah}  we obtain the following relation
% and recalling that we are interested in $\beta=0$ limit
\begin{eqnarray}
\frac{\partial\lambda^{\alpha}}{\partial\beta}&=&\frac{\partial\langle H\rangle}{\partial m^{\alpha}}=\frac{\partial^2(\beta G(\beta))}{\partial m^{\alpha}\partial\beta},\label{betah1}
\end{eqnarray}
%which is the definition of the local mean-field i.e. $h_i^{\mu}\equiv\sum_{j;\nu}J_{ij}^{\mu\nu}m_j^{\nu}$.
 From Eq. (\ref{betah1}), one can write 
\begin{eqnarray}
\frac{\partial^n \lambda^{\alpha}}{\partial\beta^n}&=&\frac{\partial^{n+1}(\beta G(\beta))}{\partial m^{\alpha}\partial\beta^n}. \label{betah2}
\end{eqnarray}
Using  Eq. (\ref{Tdef}), $\langle U^2T_2 \rangle$ can therefore be written as:
\begin{widetext}
\begin{equation}
\label{U2T2}
\langle U^2T_2 \rangle=\frac{1}{2^2}\sum_{i_1j_1}\sum_{i_2j_2}\sum_k J^{\alpha_1\beta_1}_{i_1j_1}J^{\alpha_2\beta_2}_{i_2j_2}\frac{\partial^2 \lambda_k^{\alpha_3}}{\partial\beta^2} \langle\delta S_{i_1}^{\alpha_1}\delta S_{j_1}^{\beta_1}\delta S_{i_2}^{\alpha_2}\delta S_{j_2}^{\beta_2}\delta S_{k}^{\alpha_3}\rangle.
\end{equation}
\end{widetext}
Using Eq. (\ref{betah2}), we have
\begin{equation}
\frac{\partial^2 \lambda_k^{\alpha_3}}{\partial\beta^2}=\frac{\partial^{3}(\beta G(\beta))}{\partial m^{\alpha}\partial\beta^2}=-\frac{\partial \langle U^2\rangle}{\partial m^{\alpha_3}}.
\end{equation}
The term $\langle\delta S_{i_1}^{\alpha_1}\delta S_{j_1}^{\beta_1}\delta S_{i_2}^{\alpha_2}\delta S_{j_2}^{\beta_2}\delta S_{k}^{\alpha_3}\rangle$ in Eq. (\ref{U2T2}) can be dealt with as described previously in Appendix \ref{deriv}. The outcome reads
\begin{equation}
\label{U2T2-1part}
 \langle\delta S_{i_1}^{\alpha_1}\delta S_{j_1}^{\beta_1}\delta S_{i_2}^{\alpha_2}\delta S_{j_2}^{\beta_2}\delta S_{k}^{\alpha_3}\rangle=4 \langle\delta S_{i_1}^{\alpha_1}\delta S_{i_1}^{\alpha_2}\rangle \langle\delta S_{k}^{\alpha_3}\delta S_{k}^{\beta_1}\delta S_{k}^{\beta_2}\rangle,
 \end{equation}
where the factor of 4 comes from the number of different ways of pairing site indices yielding non-vanishing results. On the other hand, 
\begin{equation}
\langle U^2 \rangle = \frac{1}{2}\sum_{ij}J_{ij}^{\alpha_1\beta_1}J_{ij}^{\alpha_2\beta_2}\langle\delta S_i^{\alpha_1}\delta S_i^{\alpha_2}\rangle\langle\delta S_j^{\beta_1}\delta S_j^{\beta_2}\rangle,
\end{equation}
and in turn
\begin{equation}
\label{U2T2-2part}
\frac{\partial\langle U^2 \rangle}{\partial m_k^{\alpha_3}} = \frac{\partial \lambda_k^{\alpha_3}}{\partial m_k^{\alpha_3}}\sum_{ij}J_{ij}^{\alpha_1\beta_1}J_{ij}^{\alpha_2\beta_2}\frac{\partial\langle\delta S_i^{\alpha_1}\delta S_i^{\alpha_2}\rangle}{\partial \lambda_k^{\alpha_3}}\langle\delta S_j^{\beta_1}\delta S_j^{\beta_2}\rangle,
\end{equation}
where $\frac{\partial \lambda_k^{\alpha_3}}{\partial m_k^{\alpha_3}}\simeq 3$. Here, we ignored higher order terms which do not contribute to the degeneracy lifting terms of fourth or sixth order in components of $\bm m$ in Eq. (\ref{U2T2}). The $\frac{\partial\langle\delta S_i^{\alpha_1}\delta S_i^{\alpha_2}\rangle}{\partial \lambda_k^{\alpha_3}}$ expression can be calculated using Eq. (\ref{chi2}). Substituting Eqs. (\ref{U2T2-1part}, \ref{U2T2-2part}) in Eq. (\ref{U2T2}), we obtain the final expression which can be represented by  a ``fused'' diagram shown in Fig. \ref{graph6}. This diagram has a new type of vertex represented with a red square. This vertex is labeled with only one site index and its mathematical expression is:
\begin{equation}
 \langle\delta S_{k}^{\alpha_3}\delta S_{k}^{\beta_1}\delta S_{k}^{\beta_2}\rangle \langle\delta S_{k}^{\alpha_3}\delta S_{k}^{\gamma_1}\delta S_{k}^{\gamma_2}\rangle,
 \end{equation}
 where again there is a sum over repeated Greek superscripts.
As a result, the final expression reads:
\begin{widetext}
\begin{equation}
\label{U2T2-f}
\langle U^2T_2 \rangle=-\frac{\partial \lambda_k^{\alpha_3}}{\partial m_k^{\alpha_3}}\sum_{ijk}J^{\alpha_1\beta_1}_{ik}J^{\alpha_2\beta_2}_{ik} J^{\alpha_3\beta_3}_{jk}J^{\alpha_4\beta_4}_{jk} \langle\delta S_i^{\alpha_1}\delta S_i^{\alpha_2}\rangle\langle\delta S_{k}^{\alpha_3}\delta S_{k}^{\beta_1}\delta S_{k}^{\beta_2}\rangle \langle\delta S_{k}^{\alpha_3}\delta S_{k}^{\beta_3}\delta S_{k}^{\beta_4}\rangle\langle\delta S_j^{\alpha_3}\delta S_j^{\alpha_4}\rangle.
\end{equation}
\end{widetext}
From this point on, one can adopt the procedure presented previously for the $\langle U^n\rangle$ terms to obtain the final result.
All other averages that contain $T_n$ e.g. $\langle U^{n_1}T_{m_1} T_{m_2}\rangle$ and $\langle T^{m_1}_{n_1} T^{m_2}_{n_2}\rangle$ 
where $n_i$, $m_i$ with $i=1,2$ are natural numbers, can be calculated following a similar procedure.

\subsection{On Diagrams} \label{sec:diagrams}
In this subsection, we make a few comments about the different diagrams that appear in the E-TAP expansion. 

First we focus on diagrams corresponding to $\langle U^n \rangle$ with $n=2,3, \cdots$, in Eq. (\ref{TAPP}). The diagrams for $n=3,4$ are illustrated in Fig. \ref{tab:graph}. In this figure, blue circles represent a given lattice site and the number of lines (bonds) connected to the circles is the number of paired $\delta S$'s at that lattice site. The degree of the vertices is indicated above each diagram and is written in the form of $(\alpha_1, \cdots, \alpha_m)$ where $m$ is the number of vertices and $\alpha_i$ is the number of bonds connected to vertex $i$. The number of times that each diagram appears in the process of pairing $\delta S$'s, referred to as diagram count, is indicated in parentheses to the right of each diagram in Fig. \ref{tab:graph}. The diagrams are distinguished by the degree of their vertices and their adjacency matrix eigenvalues. \cite{oitmaa} The adjacency matrix $M$ is a $m\times m$ matrix with matrix elements $M_{ij}$ ($0<i,j\leq m$) corresponding to the number of lines connecting the $i$-th vertex to the $j$-th vertex. We enumerate these diagrams using a computer program in which we generate all possible diagrams given a certain number of vertices of a given degree. Then, we remove all diagrams that include any onsite interactions which are forbidden since there is no onsite interaction in the Hamiltonian of Eq. (\ref{eq:hami}). For the remaining diagrams, we build their adjacency matrix which we then diagonalize. $M$ is a symmetric matrix and thus has a set of real eigenvalues which defines a graph equivalence class. Graph with the same adjacency matrix eigenvalues correspond to graphs with the same topology, i.e. they are isomorphic. \cite{diestel2000graph} The number of graphs with a given topology is noted in parentheses to the right of the graph in Fig. \ref{tab:graph}.

The fifth and sixth order terms of the E-TAP expansion in $\beta$, contain, respectively, 9 and 26 types of connected diagrams. We note that the contribution of disconnected diagrams in Eq. (\ref{TAPP}) adds up to zero.\cite{oitmaa}

The so-called fused diagrams appear in terms of the form $\langle U^mT_n\rangle$ and $\langle T_m T_n \rangle$, where $m, n \geq 2$  are positive integers. They are constructed by fusing the diagrams similar to the ones in Fig. \ref{tab:graph} together. An example of such a fused diagram and the corresponding details of its definition is given in the caption of Fig \ref{graph6}.

We note that some of the diagrams do not cover lattice sites beyond one tetrahedron, for example 
the  triangular diagram and the two site diagram in the top row of Fig. \ref{tab:graph}. 
Only, diagrams of this type contribute to fluctuations incorporated in the cluster mean-field theory (c-MFT) calculations.
However, their effect is incorporated to all orders in $\beta$ in the c-MFT calculation.

%=========================================================================================
%========================================================================================
\bibliography{ref4}
\end{document}